\documentclass[a4paper,fleqn,usenatbib]{mnras}

\usepackage{newtxtext,newtxmath,amsmath}

\usepackage[T1]{fontenc}
\usepackage{ae,aecompl}

\usepackage{graphicx}	
\usepackage{amsmath}	
\usepackage{amssymb}	

\title{Inferring AGN jet parameters using Bayesian analysis of VLBI data with non-uniform jet model}

\author[]{\parbox{\textwidth}{Ilya~N.~Pashchenko$^{1}$\thanks{E-mail: in4pashchenko@gmail.com},
Alexander V.~Plavin$^{1,2}$
}\vspace{0.4cm}\\
\parbox{\textwidth}{$^{1}$Astro Space Center of Lebedev Physical Institute, Profsoyuznaya~St.~84/32, 117997~Moscow, Russia\\
$^{2}$Moscow Institute of Physics and Technology, Dolgoprudny, Institutsky per., 9, Moscow region, 141700, Russia\\
}}

\date{Accepted 13 June 2019. Received 16 April 2019}

\pubyear{2019}

\begin{document}
\label{firstpage}
\pagerange{\pageref{firstpage}--\pageref{lastpage}}
\maketitle

\begin{abstract}

  {Physical parameters of AGN jets observed with Very Long Baseline Interferometry (VLBI) are usually inferred from the core shift measurements or flux and size measured at a peak frequency of the synchrotron spectrum. Both are preceded by modelling of the observed VLBI jet structure with a simple Gaussian templates.} 
  {We propose to infer the jets parameters using the inhomogeneous jet model directly -- bypassing the modelling of the source structure with a Gaussian templates or image deconvolution.}
  {We applied Bayesian analysis to multi-frequency VLBA observations of radio galaxy NGC~315 and found that its parsec-scale jet is well described by the inhomogeneous conical model. Our results favour electron-positron jet. We also detected a component in a counter jet. Its position implies the presence of an external absorber with a steep density gradient at close ($r=0.1$ pc) distance from the central engine.}
\end{abstract}

\begin{keywords}
methods: data analysis -- methods: statistical -- techniques: interferometric -- galaxies: jets -- radio continuum: galaxies
\end{keywords}


\section{Introduction}
\label{sec:bkmodel}
Jets observed in AGNs with Very Long Baseline Interferometery (VLBI) typically have one-sided structure with an optically thick base (called the VLBI core) that has a flat or even inverted spectrum. The position of core was found to be changing with frequency \citep{1984ApJ...276...56M}. The most plausible explanation is that the core at a given frequency is the surface of the unit optical depth to synchrotron self-absorption ($\tau_{\rm ssa}\approx1$) as predicted by the non-uniform jet model \citep{1979ApJ...232...34B, 1981ApJ...243..700K}. However, other explanations do exist \citep[e.g. standing shock,][]{2009ASPC..402..194M}. The non-uniform (or inhomogeneous) jet model was initially invoked to explain flat spectra and variability of radio sources. It assumes conical relativistically moving jet with bulk motion Lorenz factor $\Gamma$ and half-opening angle $\phi$ observed at viewing angle $\theta$. Magnetic field (assumed to be tangled) and amplitude of the particle density $K$ (both measured in a plasma rest frame) depend on the distance from cone apex as $K = K_{\rm 1}(r/r_{\rm 1})^{-n}$ and $B = B_{\rm 1}(r/r_{\rm 1})^{-m}$, where $r_{\rm 1} = 1$ pc, $B_{\rm 1}$ and $K_{\rm 1}$ are the values at the $r_{\rm 1}$. Energy distribution of particles in the plasma rest frame is assumed to be a power law $N(\gamma) = K \gamma^{-s}$ where $\gamma$ is the particle Lorenz factor. Such steady state distribution along the jet neglects effects of the radiating cooling on the shape of distribution. This could correspond to constant re-acceleration process acting across the jet volume \citep{1981ApJ...243..700K}. At the same time radially decreasing magnetic field can affect the radiative cooling rate of emitting particles at some part of the jet. This could lead to $n\approx1$ \citep{2013MNRAS.434.2380M} that is different from the canonical value $n=2$ expected from the constant speed conical isothermal jet \citep{1979ApJ...232...34B}. Exponent $s$ is related to the optically thin spectral index $\alpha$ as $\alpha = (s-1)/2$ taking the following definition of the spectral flux density $S_\nu \approx \nu^{-\alpha}$. One of the most successive applications of non-uniform model is the explanation of the aforementioned effect of the core position depending on the observing frequency --- the \textit{core shift} effect \citep[e.g.][]{1998A&A...330...79L,sokolovsky}. 

The brightness temperature distribution in a general case of arbitrary $n$, $m$ and $\alpha$ is (see Appendix~\ref{sec:extendingBK} for details):

\begin{equation}
\label{eq:Tb}
T(r_{\rm obs}, d) = \frac{c^2 C_{1}(\alpha)D^{0.5}\nu^{0.5}_{\rm obs}r_{\rm obs}^{0.5m}}{8\pi k_B C_2(\alpha) (1+z)^{0.5} B_1^{0.5}r_1^{0.5m}(\sin{\theta})^{0.5m}}(1 - e^{-\tau(r_{\rm obs}, d)})
\end{equation}
where the optical depth is:
\begin{multline}
\label{eq:tau}
\tau(r_{\rm obs}, d) =  C_{\rm 2}(\alpha)(1+z)^{-(\alpha+2.5)}D^{1.5+\alpha}K_1 B_1^{1.5+\alpha}\nu_{\mathrm{obs}}^{-(2.5+\alpha)} \\
\times 2r^{n+m(1.5+\alpha)}_{1}\left(\sin{\theta}\right)^{n+m(1.5+\alpha)-1} \sqrt{\phi_{\rm app}^2-\left(\frac{d}{r_{\rm obs}}\right)^2}r_{\rm obs}^{-(n+m(1.5+\alpha)-1)}
\end{multline}
Here the position in a sky is described by $r_{\rm obs}$ and $d$ --- distances along and perpendicular to the jet direction, $D$ -- Doppler factor, $\phi_\mathrm{app}$ is the observed jet half-opening angle, $z$ -- redshift, $C_{1}(\alpha)$ and $C_{2}(\alpha)$ -- functions of optically thin spectral index $\alpha$, $c$ -- speed of light and $k_{\rm B}$ -- Boltzmann constant.

Couple of additional assumptions were employed to derive these relations \citep{1979ApJ...232...34B} (see also Appendix~\ref{sec:deriveBK}). The first one is purely geometrical approximation of the ray path length through the jet. The second is that the absorption coefficient  in the jet is assumed to be constant along the line of sight. Both approximations are reasonable when the ratio of jet viewing angle to jet half-opening angle $\theta/\phi \sim \phi_\mathrm{app}$ is small.

\cite{1998A&A...330...79L} have shown that the inhomogeneous jet model can be used to estimate the physical parameters of the jets, e.g. magnetic field and particle density, using the measured core shift.
However the inference based on the inhomogeneous model needs some further assumptions because core shift measurements alone do not constrain the model parameters. The most widely used is the assumption of equipartition between particle and magnetic field energy densities \citep{1998A&A...330...79L}. Most of the magnetic field estimates using core shift measurements are made under this assumption \citep{1998A&A...330...79L,2009MNRAS.400...26O,sokolovsky,2012A&A...545A.113P}. \cite{2015MNRAS.451..927Z} used optically thick approximation to avoid the equipartition assumption and to tie particles number density with the observed flux density \citep[for a similar approach see also][]{2017MNRAS.468.2372N}. They found that magnetic field estimates obtained without equipartition assumption significantly differ from corresponding equipartition estimates. Both papers used fixed $s=2$, that corresponds to optically thin spectral index $\alpha=0.5$. Considering strong dependence of the model predictions on this particular parameter (Appendix~\ref{sec:extendingBK}) it is desirable to simultaneously infer it from the observed data. At the same time even these simplified approaches need measurements of the core shifts. However core shift estimates obtained using the fitting of the VLBI core with a Gaussian are biased with bias that depends on resolution \citep{2019MNRAS.485.1822P} and observed jet parameters (Pashchenko et al., in prep.).

\cite{finke} used analytical expressions for flux density and core shift of the \cite{1979ApJ...232...34B} model to estimate physical jets properties. They used fixed exponents of the magnetic field and particle density radial dependence and used jet power as an additional observational constrain. The resulting estimates have large uncertainties due to low accuracy of the available core shift estimates and unknown value of the particle energy spectrum exponent $s$, which is poorly constrained by the data they used.
\cite{2019arXiv190400106F} used special-relativistic hydrodynamic (SRHD) simulations as a jet model to fit multifrequency VLBI observations of the radio galaxy NGC~1052. They used images stacked over several epochs to smooth out numerous inhomogeneities travelling along the flow and minimised the difference between the observed stacked and model images at several frequencies. Although they used artificially created VLBI images obtained with the same VLBI array and visibility noise as in the observed data, comparing in the image plane -- where the noise is correlated -- complicates the uncertainty analysis of the fit. 
Furthermore, the optimization approach could suffer from the local minima and miss separated high probability parameter regions. 

In this paper we use the inhomogeneous jet model to make inference from the radio interferometric visibility data directly -- bypassing the modelling of the source structure with a Gaussian templates or image deconvolution.

Throughout the paper, we adopt the $\Lambda$CDM cosmology with $\Omega_m=0.287$, $\Omega_\Lambda=0.7185$ and $H_0=69.3$~km~s$^{-1}$~Mpc$^{-1}$ \citep{2013ApJS..208...19H}.

\section{Data}
\label{sec:data}

We searched for the source with a featureless straight jet and not small jet viewing angle. The first requirement is needed because the model of \cite{1979ApJ...232...34B} does not account for inhomogeneities in a jet, e.g. shocks. However this is not a principal obstacle for using our approach as shocks can be described as additional components over the top of the smooth underlying jet. 
The model we propose does not handle jet curvature as-is, so for now we need to limit our selection to straight jets. The last geometric requirement of relatively high jet viewing angle implies that the chosen object should be a radio galaxy \citep{1995PASP..107..803U}. It is due to one of the  approximations in the original paper and can be completely abandoned by numerically solving corresponding radiative transfer instead of using the analytical solution (\ref{eq:Tb}).

\begin{figure}
   \centering
   \includegraphics[width=\columnwidth, trim=0.3cm 0.5cm 0.3cm 0.3cm]{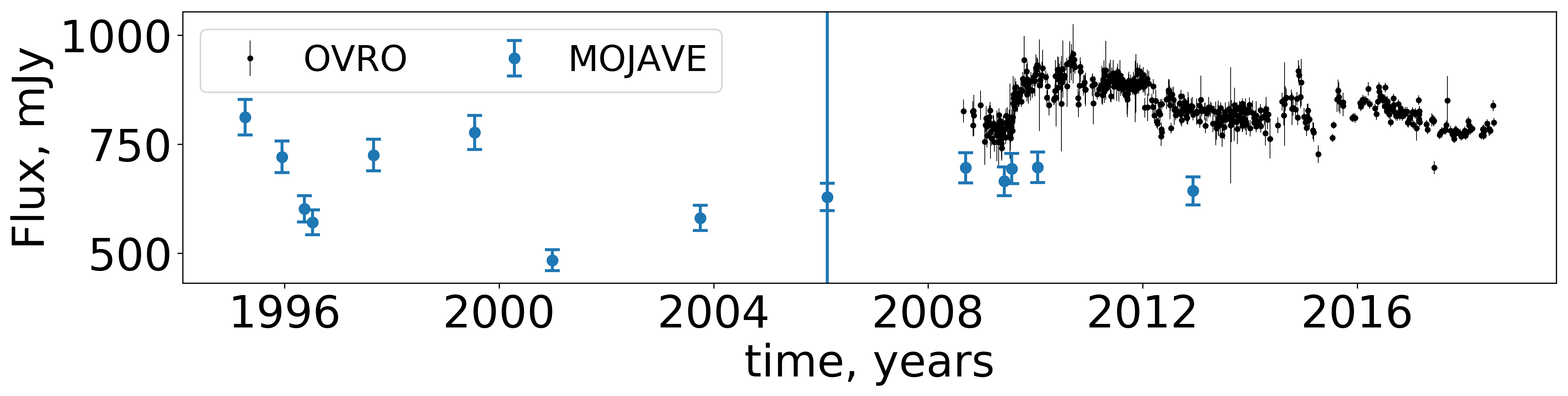}
   \caption{Single-dish 15 GHz OVRO light curve and MOJAVE VLBA-scale fluxes. Uncertainties of the MOJAVE VLBA fluxes are 5\% according to \protect\cite{2005AJ....130.2473K}. Vertical line shows MOJAVE epoch with multifrequency data used in our analysis.}
   \label{fig:lightcurve}
\end{figure}

We choose radio source NGC~315 \citep{1976Natur.262..179B} which is a close ($z=0.0165$) giant FR~I \citep{1974MNRAS.167P..31F} radio galaxy. It resides in elliptical galaxy and optically classified as broad-lined LINER \citep{1997ApJS..112..391H}.
Flux density monitoring of NGC~315 with \texttt{Westerbork} synthesis radio telescope at 5 GHz from 1974 to 1980 revealed constant flux \citep{1983A&A...120..297E}. \cite{1999ApJ...519..108C} noted the flare that was observed using VLA in 1990--1995 that increased the arcsecond core flux 1.5x \citep[from 588 mJy in 1990 Sep to 746 mJy in 1994 June,][]{1993ApJ...408...81V,1999ApJ...519..108C}. 
OVRO \citep{2011ApJS..194...29R} 15 GHz light curve and MOJAVE \citep{2005AJ....130.2473K} 15.4 GHz VLBA-scale fluxes are presented in Figure~\ref{fig:lightcurve}.
\cite{2005MNRAS.363.1223C} modelled kpc-structure of this source with a relativistic decelerated jet model and estimated viewing angle $\theta = 38\pm2 \deg$ with initial speed $\beta \approx 0.9$ (in units of speed of light $c$). On epoch (2006--02--12) the source was observed at VLBA at several frequencies (15.4, 12.1, 8.4 and 8.1 GHz -- u, j, y and x bands) in MOJAVE survey \citep{2009AJ....137.3718L}. Corresponding self-calibrated interferometric visibilities were taken from the MOJAVE database\footnote{\url{http://www.physics.purdue.edu/astro/MOJAVE/sourcepages/0055+300.shtml}}.

\section{Bayesian inference and probabilistic model}
\label{sec:probmodel}

\subsection{Re-parametrization and additional assumptions}
\label{sec:reparametrization}
The model of the observed brightness distribution from (\ref{eq:Tb}) besides jet parameters $\theta$, $\phi$, $B_1$, $K_1$, $m$, $n$, $\alpha$ includes $dx$ and $dy$ -- the location of the jet apex in the image, and $\theta_{\rm rot}$ -- jet position angle on the celestial sphere.

Due to degeneracies between model parameters (e.g. $B_1$ and $K_1$, $\theta$ and $\phi$) we introduce a different parametrization (\ref{eq:TparametrizationNatural} and \ref{eq:tauparametrizationNatural}) and use the observed half-opening angle $\phi_{\rm obs}$. Thus our re-parametrized model has the following parameters: $dx$, $dy$, $\theta_{\rm rot}$, $\phi_{\rm app}$, $A_1$, $A_2$, $m$, $n$, $\alpha$. There is still an unidentifiable combination of parameters $n + m(1.5+\alpha)-1$ in the exponent of the observed position in the expression for the optical depth. This, together with $A_2$, defines the characteristic size of the core and thus the core shift effect. However the optically thin spectral index $\alpha$ can be constrained by fitting data at several frequencies simultaneously. The total flux spectrum also constrains the exponents $n$ and $m$ \citep{1981ApJ...243..700K}. However we found that this weakly bounds parameters $m, n$ and $\alpha$.
Fitting the model with arbitrary exponents $n$ and $m$ to the multifrequency data shows that the degeneracy between them still remains. This could results in poor exploration of their distribution so this item deserves additional investigation. Thus we further assume that the particles to the magnetic field energy density ratio is constant, i.e. $n = 2m$. 
This assumption is weaker than the equipartition assumption. However, it could be violated during flares \citep{1999ApJ...521..509L,2017MNRAS.468.4478L,2019MNRAS.485.1822P}, when region of e.g. higher plasma density is travelling along the jet. NGC~315 light curve presented in Figure~\ref{fig:lightcurve} shows only a slowly rising VLBA-scale flux increasing by 20\% from epoch 2001 to 2009. Thus, the proposed assumption seems reasonable.
Moreover, as shown in \citep{2014AJ....147..143H} distribution of the core spectral indices in MOJAVE sample peaks at $\alpha=0$ thus supporting the assumption of a constant particle-to-magnetic energy density ratio ($n = 2m$).

We stress that the re-parametrization (\ref{eq:TparametrizationNatural} and \ref{eq:tauparametrizationNatural}) is not necessary, i.e. its usage does not bring any new information, nor bias the estimates of the physical parameters (e.g. $B_1$ and $K_1$) in any way. One could make the inference using original model with parameters that include $B_1$ and $K_1$. But using re-parametrized model without degeneracies significantly speeds up the inference (i.e. sampling high dimensional posterior distribution of the model parameters - see Section~\ref{sec:posterior}). Also sampling of high dimensional distributions with pronounced degeneracies is still a challenge for most of the algorithms in wide usage.

As we fit the model to self-calibrated data, we assume that gains of the individual antennas are accounted for during self-calibration process \citep{2017isra.book.....T}. However the remaining amplitude scale factors should be taken into account, because they affect the spectrum and thus our inference. The corresponding absolute calibration uncertainty was estimated by \cite{2014AJ....147..143H} as $\sigma_{\rm b}=5\%$ for x, y and u-bands and 7.5$\%$ for j-band. The scale factors are degenerate as their simultaneous multiplication by the same number is equivalent to scaling $A_1$ parameter with this number (see \ref{eq:TparametrizationNatural} and \ref{eq:tauparametrizationNatural}). To break this degeneracy we put an additional constrain on amplitude scale factors: $\prod_b g_{b} = 1$. However, this leaves an uncertainty of the product not being exactly unity. The remaining uncertainty due to scaling each of the $g_{b}$ to the same factor can be accounted for using priors for $g_{b}$ after the sampling (see Appendix~\ref{sec:accountmeangain} for details of our implementation). Another option is to fix one amplitude scale factor to 1 \citep{2017MNRAS.464.4306N}. However this makes it difficult to account for the uncertainty of other model parameters due to this chosen scale factor not being exactly 1.

Finally, our preliminary fits revealed a significant component in a counter-jet side that is fundamentally not accounted by the model described above. Emission on the counter-jet side was detected by examining the residuals in terms of visibilities and reconstructed CLEAN maps. To account for this emission we extended our model by including a point-like component located on the jet axis at some distance from the apex. We used a point component because the sizes of the circular Gaussian components if used are much less than a beam size and weakly constrained. We treat this as component being unresolved. Thus our extended model has 25 parameters in total: 6 non-uniform jet model parameters, jet position angle, location of the jet apex (2 for each of 4 bands), 3 parameters of the amplitude scale and 8 parameters of the counter-jet component (2 for each band).

\subsection{Bayesian inference and sampling}
\label{sec:posterior}
Posterior distribution of the model parameters $\bold{\theta}$ is expressed as the product of the likelihood and prior distribution \citep[Bayes Theorem,][]{bayes,laplace}:
\begin{equation}
    P(\bold{\theta}| \bold{V}_{\rm obs}) \sim P(\bold{V}_{\rm obs} | \bold{\theta}) \cdot P_{\rm pr}(\bold{\theta})
\label{eq:bayes_theorem}
\end{equation}
where $\bold{V}_{\rm obs}$ -- the observed data and the prior distribution is itself a product of priors for individuals model parameters:
\begin{equation}
    P_{\rm pr}(\bold{\theta}) = \prod_{i=1}^{i=N_{\rm params}} P_{\rm pr}(\theta_{i}).
\label{eq:prior}
\end{equation}

We fit the model to the self-calibrated visibility and assume the gaussian noise in its real and imaginary parts.
Thus, the likelihood:
\begin{equation}
\begin{split}
    P(\bold{V}_{\rm obs} | \bold{\theta}) = \prod_{b=X,Y,J,U} \prod_{i=1}^{N_{{\rm vis},b}} \frac{1}{2 \pi \sigma^2_{b,i}} \\
    \exp\left(-\frac{|g_b V_{{\rm model},b,i}(\bold{\theta}) - V_{{\rm obs},b,i}|^2}{2 \cdot \sigma^2_{b,bl(i)}}\right)
\label{eq:vismodellik}
\end{split}
\end{equation}
where $g_b$ -- amplitude scale factor for frequency band $b$, constrained as described in Section~\ref{sec:reparametrization}, $V_{{\rm model},b,i}(\bold{\theta})$ -- Fourier transform of a model at frequency band $b$ with parameter vector $\bold{\theta}$ at $i$-th ($u$, $v$)-point, $V_{{\rm obs}, b,i}$ -- observed visibility function at band $b$ at $i$-th (u, v)-point, $\bold{V}_{{\rm obs}, b} = (V_{{\rm obs}, b,i}), i=\overline{1, N_{{\rm vis}, b}}$ -- vector of the observed visibilities at frequency band $b$, $\sigma^2_{b,bl(i)}$ -- dispersion of normal noise on $bl(i)$-th baseline for real/imaginary part of the Stokes I visibility\footnote{Stokes I visibilities are calculated as half sum of parallel hand correlations, i.e. $0.5\cdot(<RR*> + <LL*>)$ where L and R are voltages from left and right circular polarized feeds, angular brackets means correlation and star denotes complex conjugation \citep{2017isra.book.....T}} at frequency band $b$. We estimated them with the successive differences approach \citep{briggs} for each baseline.

We used non-uniform pixel size for calculating the model image with pixel size increasing from 1 $\mu$as close to the jet apex to 0.1 mas at 20~mas along the jet. This is justified by the fact that the model we use has smooth structure at large scales and more compact details in the core region. Also this significantly speeds up calculation of the model image and Fourier Transform to the visibility space, compared to using the same small pixel size for the whole jet.

We used normal priors for amplitude scale factors $g_b$, positions $dx_b, dy_b$, angles $\phi_\mathrm{app}, \theta_\mathrm{rot}$, uniform for positions of the counter-jet component $r_b$ and lognormal for $m$, $\alpha$, $A_1, A_2$, fluxes of the counter-jet component $F_b$ and checked that our sampled posterior is not bounded by the corresponding priors. We used wide uninformative priors for amplitude scale factors $g_b \sim N(1, 0.1)$ in sampling and more informed priors $N(1, \sigma_{\rm b})$ in post-sampling step of accounting for the error due to unknown mean scale factor (Appendix~\ref{sec:accountmeangain}).

The sampling from posterior distribution was done using a \textit{nested sampling} \citep{doi:10.1063/1.1835238} algorithm as implemented in \texttt{\mbox{PoLyChord}} \citep{polychord1,polychord2}. It implements constrained sampling from prior distribution using slice sampling and allows effectively sample multimodal high dimensional posterior distributions with linear degeneracies.
The resulting posterior distribution of the parameters is presented in Figure~\ref{fig:corner}. Position of the cone apex and parameters of counter-jet component are only shown for 15.4 and 8.1 GHz for compactness.
Here the prior distributions of individual parameters are plotted as the orange lines on diagonal plots together with histograms of the marginalized posterior distributions. 

Finally, we note that contrary to fitting in the image plane \citep[e.g.][]{2019arXiv190400106F}, ($u$, $v$)-plane fitting allows to make inference even if the image deconvolution is not feasible (e.g. in Space VLBI observations with sparse coverage of the ($u$, $v$)-plane). Moreover, sampling the full multidimensional posterior distribution of the model parameters ensures to detect possible high-probability modes of the posterior that could be missed by optimisation algorithms. This also provides justified uncertainty estimates of the model parameters.

\section{Model evaluation}
\label{sec:modelevaluation}

\subsection{Comparing with the observed visibilities and residuals images}
\label{sec:residuals}

\begin{figure*}
   \centering
   \includegraphics[width=\textwidth]{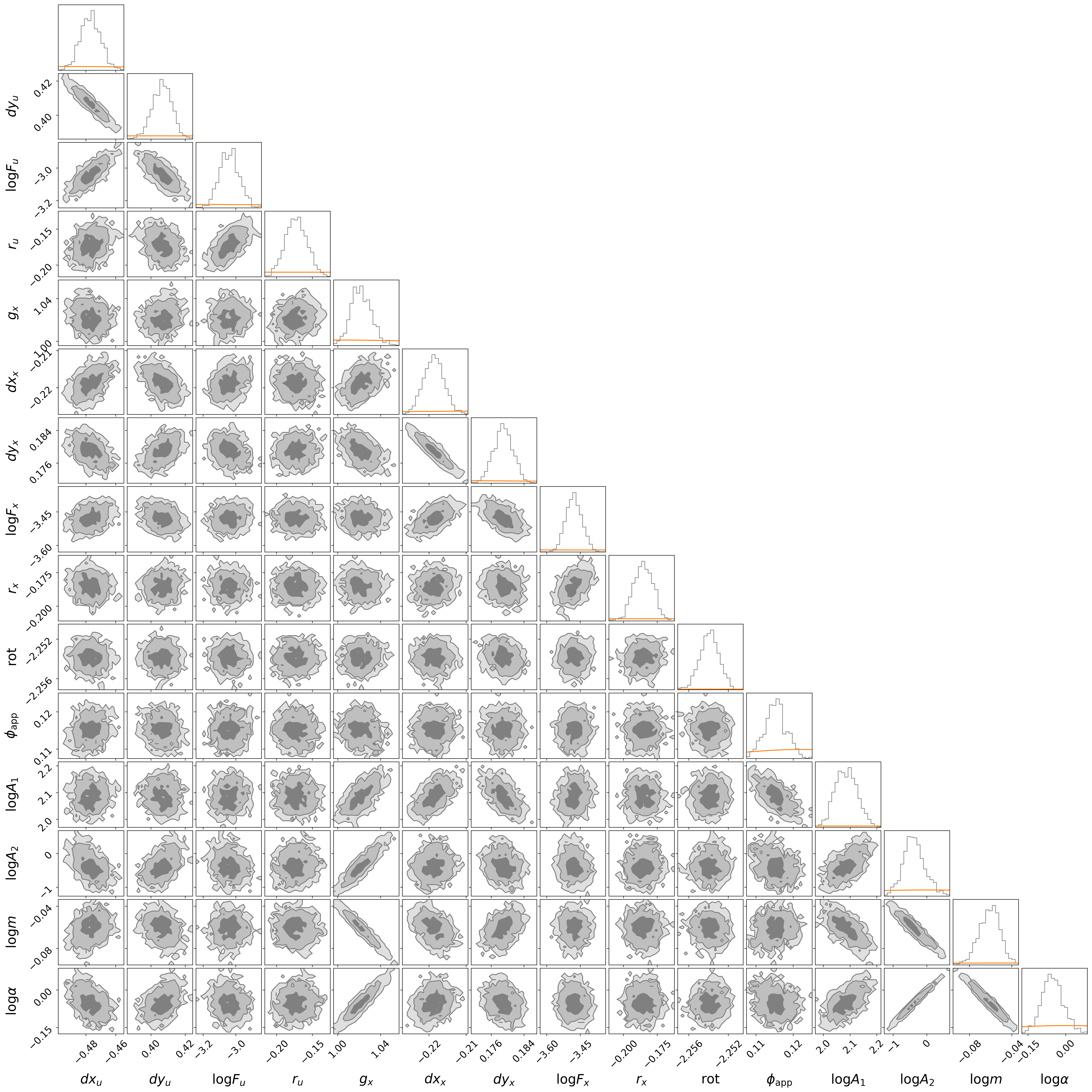}
      \caption{Posterior distribution of the parameters. Position of the cone apex ($dx$, $dy$ in mas) and parameters of counter-jet component (logarithm of the flux $\log{F}$ in Jy and distance from jet apex $r$ in mas) are shown for 15.4 and 8.1 GHz ($u$ and $x$-bands) only. The diagonal shows marginalized distributions of each parameter and the orange curves show priors of the corresponding parameters. Other plots show joint distribution of pairs of the parameters marginalized over all other parameters. Different shades of gray show 1, 2 and 3 $\sigma$ levels. }
         \label{fig:corner}
   \end{figure*}

\begin{figure}
   \centering
   \includegraphics[width=\columnwidth, trim=0.3cm 0.5cm 0.3cm 0.3cm]{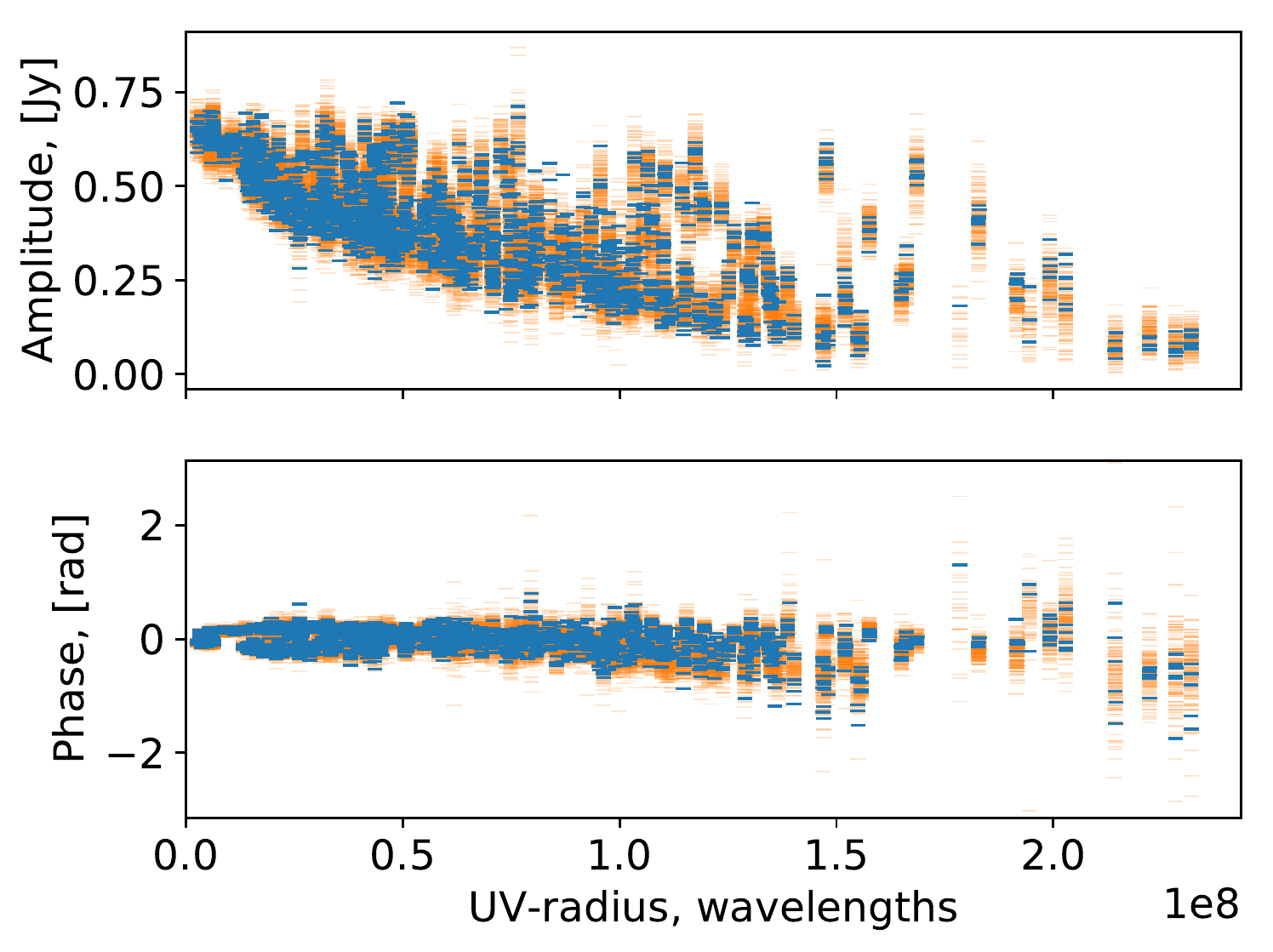}
   \includegraphics[width=\columnwidth, trim=0.3cm 0.5cm 0.3cm 0.3cm]{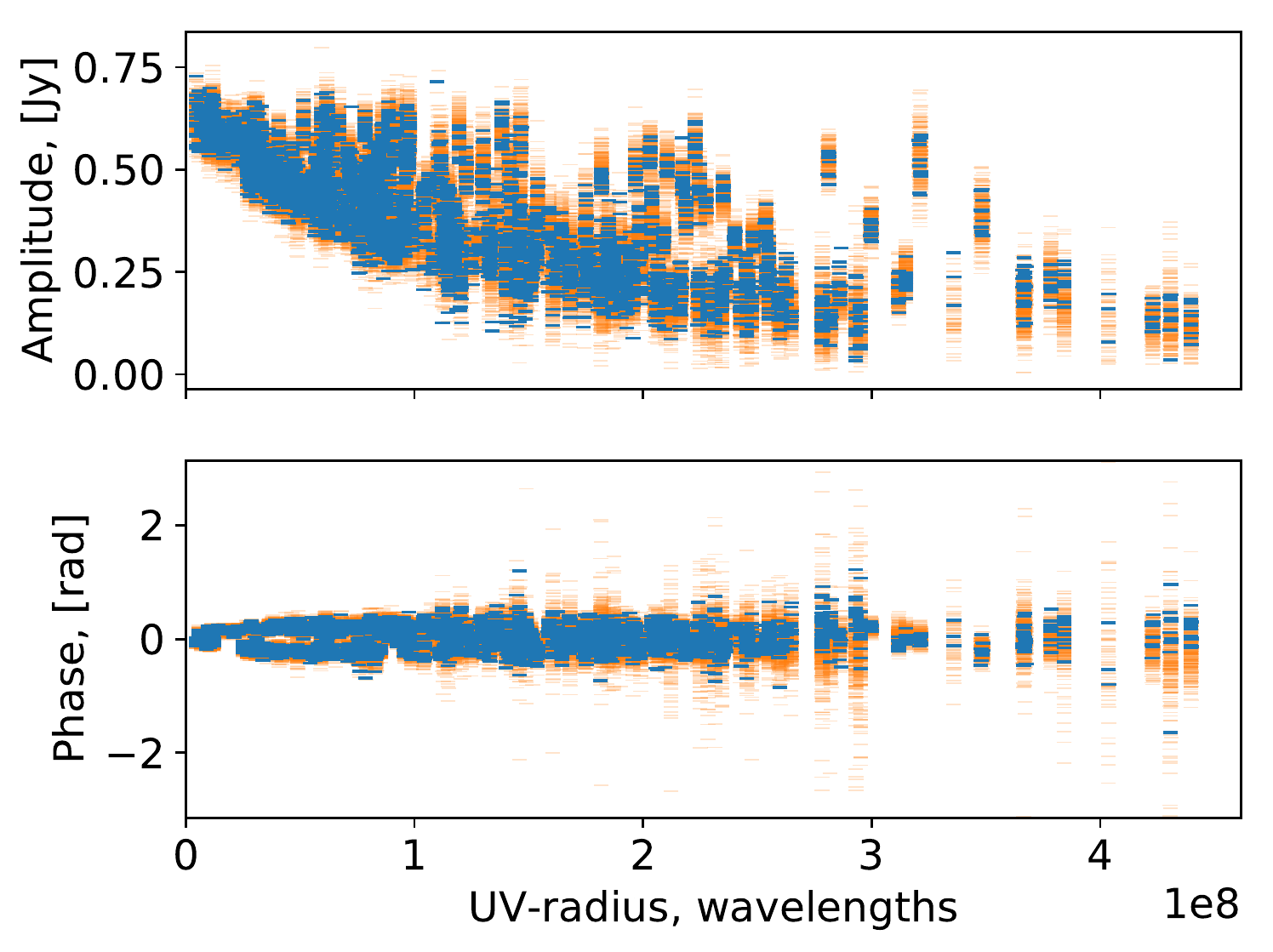}
      \caption{Plot of the observed (\textit{blue}) and model (\textit{orange}) amplitudes and phases with realistic noise added vs. radial ($u$, $v$)-distance for 8.1 and 15.4 GHz.}
       \label{fig:xuradplot}
   \end{figure}

Visibility amplitude and phase dependence on radial distance in ($u$, $v$)-plane is presented for the observed and model data for 8.1 and 15.4 GHz in Figure~\ref{fig:xuradplot}. To account for the uncertainty in estimated model parameters we plotted model visibilities not only for the single ``best'' value of the parameter vector (e.g. Maximum Likelihood) but for a sample of the parameters (24 samples) drawn from the posterior distribution (Figure~\ref{fig:corner}). We also added the observed noise to the model visibilities for each sample from the posterior to aid the comparison with the observed data. This step is necessary because noise contributes significantly to the amplitude of the observed visibilities at the largest baselines with low signal-to-noise ratio. We also accounted for the uncertainty of the mean amplitude scale not being exactly 1.

\begin{figure}
  \centering

  \includegraphics[width=0.85\columnwidth, trim=0.3cm 0.5cm 0.3cm 0.3cm]{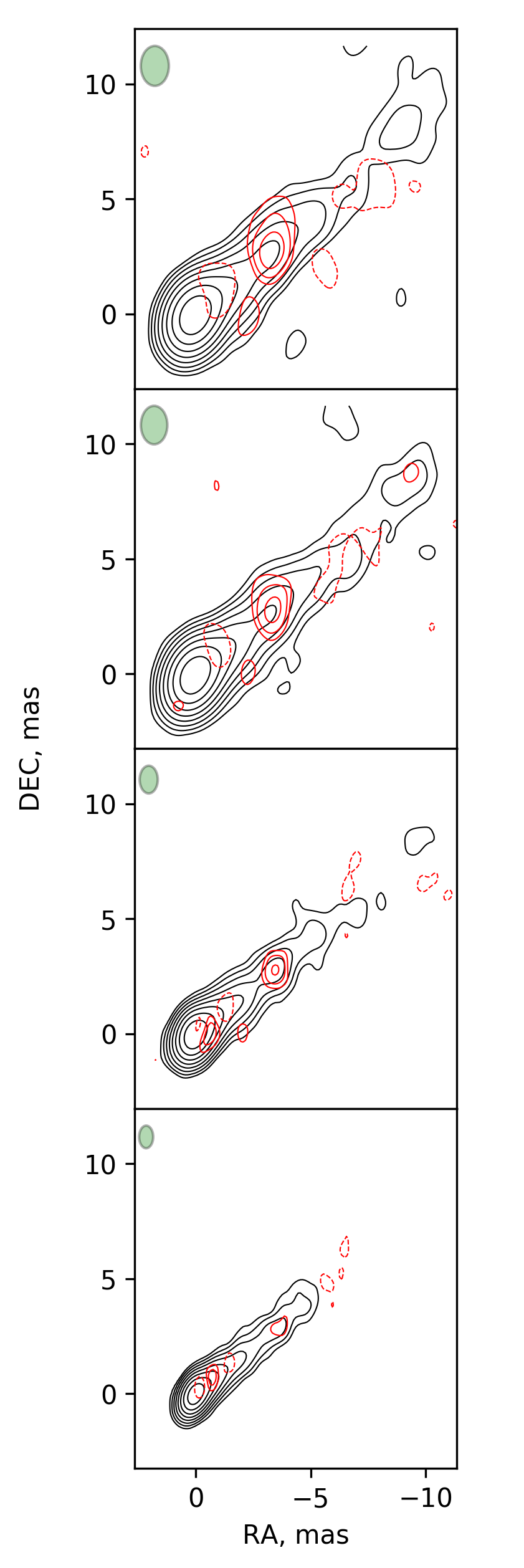}
      \caption{CLEAN images of the observed data (the black contours) and difference between the observed and best-model data including contribution from the counter-jet (the red contours) at (from up to bottom) 8.1, 8.4, 12.1 and 15.4 GHz. Minimal contour is 1.7 mJy/beam and spacing between contours is factor 2.}
      \label{fig:imdiffcj}
  \end{figure}

The original CLEAN images superimposed with the CLEAN images of the differences between the observations and our best model\footnote{We use Maximum Likelihood parameters here.} are plotted in Figure~\ref{fig:imdiffcj}. The most obvious is a residual component in the jet at DEC$\approx$4 at all bands. We discuss it in Section~\ref{sec:comp4mas} further. Also it is apparent that the model fits the inner jet region better at lower frequencies.

\subsection{Comparing with CLEAN and Gaussians models}
\label{sec:compareCLEANandGauss}

As a part of the model evaluation procedure we compared our model with a CLEAN model and a model consisting of several circular Gaussian components. CLEAN and model fitting were done in \texttt{Difmap} package \citep{1997ASPC..125...77S}. Our model has 25 parameters and describes the observed data at 4 frequency bands simultaneously. To describe the same data using model with circular Gaussian components one needs $\sim100$ parameters. Corresponding CLEAN model has $\sim$1000 parameters at each frequency band (position and flux of $\delta$--functions). Various fit statistics, including sum of the $\chi^2$, reduced chi-squared $\chi_{\rm red}^2$ over all 4 bands for CLEAN, Gaussians and our final model are presented in Table~\ref{tab:rchi2}\footnote{Note the critique of using $\chi_{\rm red}^2$ \citep{2010arXiv1012.3754A}}. Despite the model consisting of Gaussians has slightly lower $\chi^2_\mathrm{red}$, it has unclear physical interpretation considering the smooth jet structure. The inherently unphysical but highly flexible CLEAN model has the same $\chi^2_\mathrm{red}$ as the non-uniform jet model for this source.

\begin{table}
\begin{center}
\caption{Fit statistics for different models}
\label{tab:rchi2}
 \begin{tabular}{c@{~~~}c@{~~~}c@{~~~}c@{~~~}c@{~~~}c}
\hline
model & $\chi^2$ & $N_{\rm param}$ & DoF & $\chi_{\rm red}^2$\\
\hline
 CLEAN  & 25969.8 & 3537 & 22311 & 1.16$\pm$0.005\\
 Gaussians  & 27433.6 & 136 & 25712 & 1.07$\pm$0.004\\
 Non-uniform jet  & 29747.9 & 25 & 25823 & 1.15$\pm$0.004 \\

\hline
\end{tabular}
\end{center}
{\bf Column designation:}
Col.~1~-- Model,
Col.~2~-- $\chi^2$, all bands,
Col.~3~-- Number of the parameters in model, all bands,
Col.~4~-- Number of degrees of freedom (DoF), all bands,
Col.~5~-- reduced chi-squared $\chi_{\rm red}^2$ and its statistical error $\sqrt{2/{\rm DoF}}$, all bands, 
\end{table}

\subsection{Comparing with the observed data - Posterior Predictive Check}
\label{sec:comparedata}

Adequate model should create data similar to the observed data \citep{davis1995,lindsay2009}. Posterior predictive check \citep[PPC,][]{gelman2013bayesian} can be used as a measure of discrepancy that includes the uncertainty associated with the
estimated model parameters. PPC consists in simulating the data under the fitted model and then comparing these with the observed data to discover possible systematic differences.

We made the posterior predictive check for the total CLEAN model flux by sampling parameters from the posterior (Figure~\ref{fig:corner}) 200 times, each time constructing the model image, transforming it to ($u$,$v$)-plane, applying corresponding amplitude scaling factors, adding the noise estimated from the observed visibilities and CLEANing with the same parameters (image and pixel size) as for the observed data using \texttt{Difmap} package. All observed values lie in middle quartiles of the posterior predictive distributions. The corresponding percentiles of the observed CLEAN model flux among the posterior predictive distributions are 37\%, 65\%, 45\% and 64\% for frequencies 15.4, 12.1, 8.4 and 8.1 GHz correspondingly. Thus, the model describes the total flux well.

\begin{figure}
   \centering
   \includegraphics[width=\columnwidth, trim=0.3cm 0.5cm 0.3cm 0.3cm]{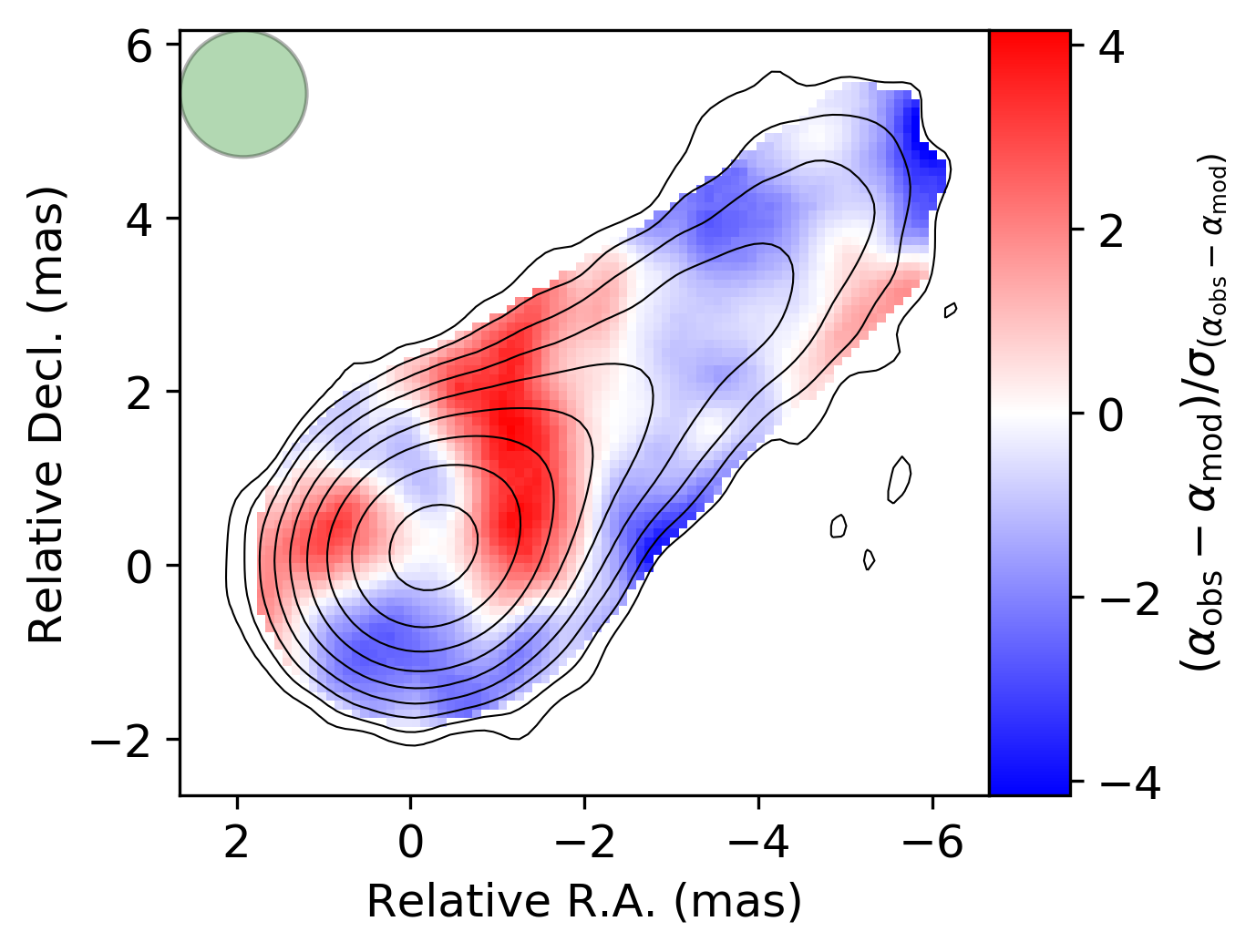}
      \caption{The difference between the observed and model spectral index normalized by the corresponding pixel standard deviation. The green circle shows the common restoring beam.}
       \label{fig:spixppcimage}
   \end{figure}

\begin{figure}
   \centering
   \includegraphics[width=\columnwidth, trim=0.3cm 0.5cm 0.3cm 0.3cm]{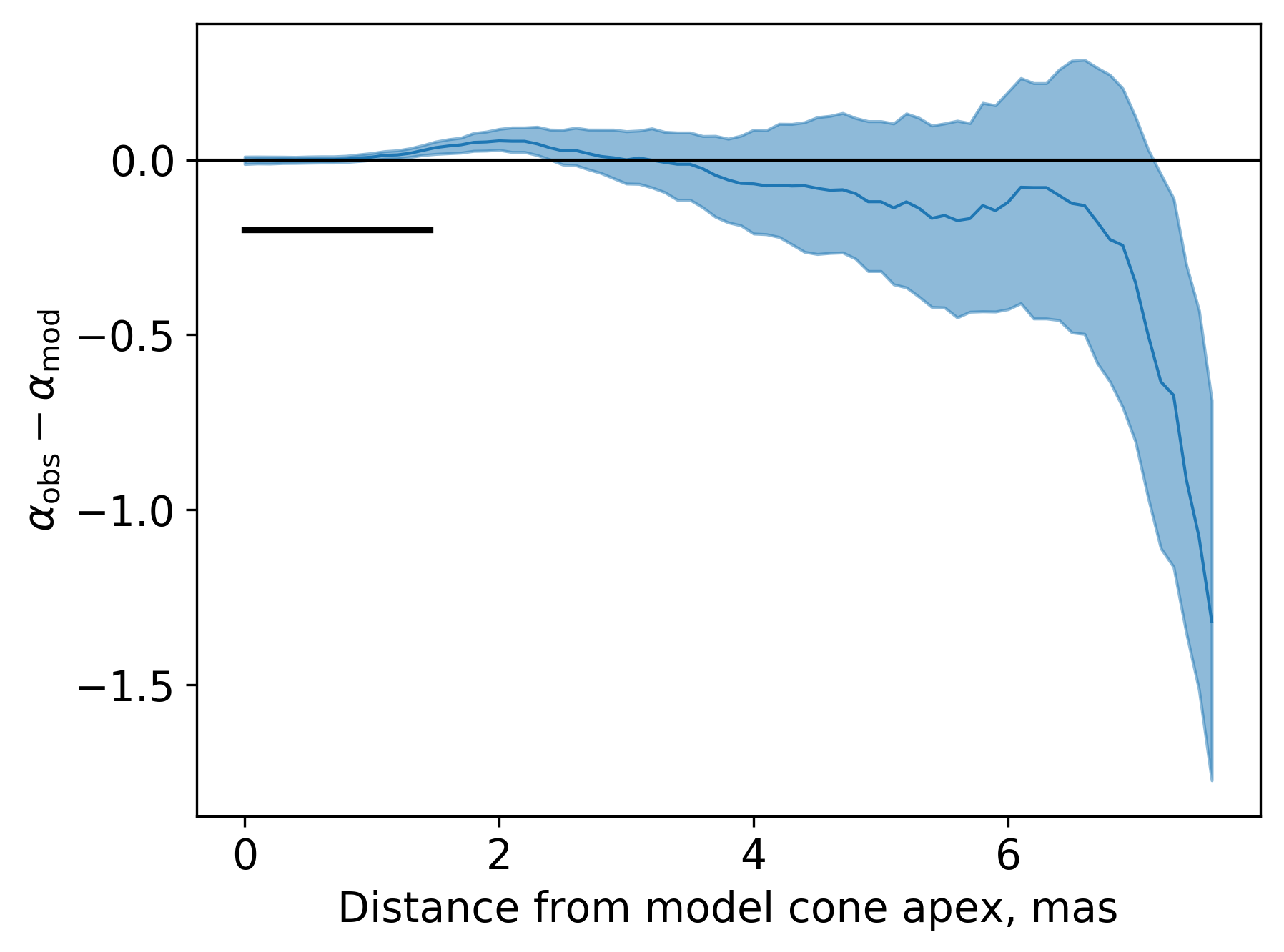}
      \caption{The slice of the difference between the observed and model spectral index along the model cone axes. Median value and 95\%--pointwise confidence band for 100 posterior samples are shown, The thick black line shows the common restoring beam.}
       \label{fig:spixppcslice}
   \end{figure}

\cite{2014AJ....147..143H} found spectral index in the jet region to be $\approx1.1$ with a typical error $\approx0.2$. Notably they used the component we interpret being on the counter-jet side as the core to align the images at different frequencies. However this should not influence the jet spectral index. Our modelling results in the median value for $\alpha = 0.95$ with 68\%-credible interval -- (0.91, 0.99), consistent with theirs.
We conducted posterior predictive check for spectral index in the following way. We chose 200 random samples from the posterior distribution of model parameters (Figure~\ref{fig:corner}). Then for each sample we simulated the observed data in the same way as for the total flux PPC. To align CLEAN images at different frequencies we shifted the phases of model visibilities and put the position of the jet apex at each frequency to the phase centre. The same shifts were also applied to the real data. After CLEANing of all data sets with the same pixel size and restoring beam we obtained many realizations of the spectral index maps for the model and observed data (conditioning on the current model parameters). The difference between the observed and model spectral index normalized by the corresponding standard deviation is presented in Figure~\ref{fig:spixppcimage}. Apparently, the model describes the observed spectral index distribution well besides the barely significant difference in the region of inner jet at $\approx2$ mas from the cone apex. Here the observed spectral index is flatter than the model one. To show the magnitude of the effect we plot the posterior distribution of the difference between the observed and model spectral index map along the model jet axis in Figure~\ref{fig:spixppcslice}. Corresponding difference in the on-axis spectral index is only $\approx$0.05.

\citep{2017MNRAS.468.4992P} obtained the median value of half-opening angle $\phi_{\rm app} = 3.45\pm0.05$ degrees for 15.4 GHz VLBA images stacked over 14 epochs, while our estimate for epoch 2006-02-12 is nearly two times larger -- median 6.61 with 95\% credible interval (6.27, 6.94). There are several possible causes of such inconsistency. \citep{2017MNRAS.468.4992P} used deconvolved FWHM of the Gaussians fitted to the transverse slices of the observed brightness distribution to calculate jet width and half-opening angle, while our estimate concerns intrinsic jet geometry. Also they possibly used the counter-jet component as a core, making $\phi_{\rm app}$ smaller and analyzed stacked image.

\begin{figure}
   \centering
   \includegraphics[width=\columnwidth, trim=0.3cm 0.5cm 0.3cm 0.3cm]{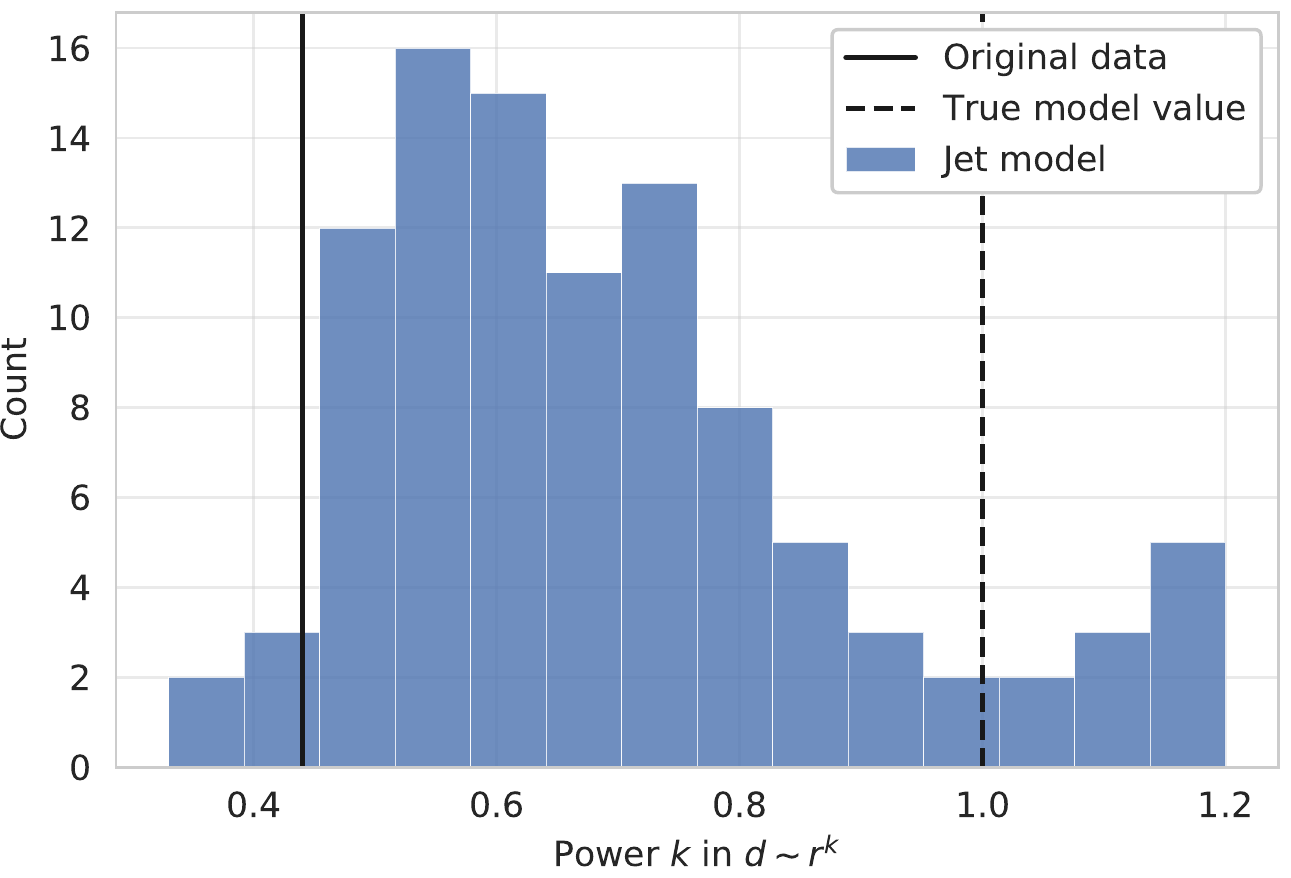}
      \caption{Distribution of the shape parameter $k$ from $R \propto r^{k}$ for 100 simulated data sets generated from the posterior together with the observed value and the intrinsic model value $k=1$.}
       \label{fig:jetshapemodel}
   \end{figure}

\citep{2017MNRAS.468.4992P} found exponent of the jet width $d$ radial dependence $k = 0.86 \pm 0.01$, where $d \propto r^{k}$. Although we can not compare to their results directly as they used an image stacked over many epochs, we conducted PPC using the jet shape at single epoch used in our analysis. We generated artificial data from the posterior distribution of model parameters (Figure~\ref{fig:corner}) and added noise as in the original data. Then we CLEANed each artificial data set and convolved corresponding CLEAN models with the same circular beam. For each of the obtained image we applied the method of \citep{2017MNRAS.468.4992P} to calculate shape parameter $k$. The posterior predictive distribution together with the observed value are presented in Figure~\ref{fig:jetshapemodel}. The width of the distribution is large but the observed value lies at its low tail (with $p$-value 0.06 for the corresponding two-sided statistical test). Interesting that the PPC distribution is well below the model value $k=1$. This controversy deserves further investigation. It could imply that the conclusions from such jet shape measurements to the outflow geometry should be treated with caution.

\section{Inferred physical properties of the jet}
\label{sec:resultsphys}

\subsection{Spectral index and particle energy distribution}
\label{sec:spixacc}

We obtained the median value of the optically thin spectral index $\alpha = 0.95$ and 
68\% credible interval -- (0.91, 0.99). This results in a power law exponent of the emitting electrons $s=2.90$ with 95\% credible interval (2.75, 3.07). It is significantly higher than $s=2.0$ assumed in the original paper \citep{1979ApJ...232...34B} and also significantly differs from both $s=2.5$ attributed to weakly magnetized shock acceleration and ultra-relativistic limit $s=2.2$ \citep{2015SSRv..191..519S}.

However in strong magnetic field the emitting particles are cooled efficiently \citep{1962SvA.....6..317K} and the break in their energy spectrum occurs at some $\gamma_{\rm br}$. Above the break $\gamma > \gamma_{\rm br}$ the power law steepens from the original (acceleration) value $s_{\rm acc}$ to $s_{\rm acc} + 1$. If we observe the steepen part of corresponding synchrotron spectrum then $s_{\rm acc} = s - 1 = 1.90$ with 95\% credible interval (1.75, 2.07). This also significantly differs from the ultra-relativistic shocks universal value.
One can estimate the break frequency $\nu_{\rm br}$ at given location $r$ by equating the jet travel time to a distance $r$ with synchrotron cooling time \citep{1979ApJ...232...34B,1981ApJ...243..700K}. We obtain the median $\nu_{\rm br} = 387$ GHz with 68\%-credible interval (217, 701) [GHz] at the position of the VLBI core at 15.4 GHz. This implies that the observed emission resides well below the break and, thus, our model is self-consistent.

\cite{1993ApJ...408...81V} observed optically thin spectral index $\alpha_{5}^{8.4} \approx 0.5$ between 5 and 8.4 GHz and sharp steepening at $r\approx 4$ mas from the core using VLBI observations at close epochs (1989 Apr -- 1990 Nov). They conclude that the break frequency $\nu_{\rm br}\approx5$ GHz at distance $r$ =  4 mas from the core. Thus at our frequency bands (i.e. above $\nu_{\rm br}$) we should see a steepen spectrum at this location. Interesting that the steepening is observed at 6 mas in our residuals between data and model (Figure~\ref{fig:spixppcslice}). However, \cite{2014AJ....147..143H} showed that for steepening with distance the jet should be collimating with radius $R \propto r^{k}$ where $k < 2/3$ assuming magnetic field with dominating transverse component. Another explanation is a cutoff in synchrotron spectrum due to high energy cutoff $\gamma_{\rm max}$ of the electron energy spectrum \citep{2014AJ....147..143H}. This could imply $\gamma_{\rm max}=370\pm20$ at this distance assuming that emission at given frequency 8.1 GHz is dominated by electrons with $\gamma_{\rm rad} \approx \sqrt{(\nu_{\rm 8.1 GHz}/D)/\nu_{B_{\rm 6 mas}}} = 370$, where $\nu_{B}$ - Larmor frequency for magnetic field at a given radial distance.

\subsection{Magnetic field and particle density}
\label{sec:fields}

We obtained values of the exponents $m$ and $n$, that determine the radial dependence of the magnetic field and particle number density, close to the canonical \citep{1979ApJ...232...34B}: $m = 0.94 \pm 0.01$ and $n = 1.88\pm0.02$ (68\% credible intervals). The value $m\approx1$ implies the dominant transverse component of the magnetic field.

\begin{figure}
   \centering
   \includegraphics[width=\columnwidth, trim=0.3cm 0.5cm 0.3cm 0.3cm]{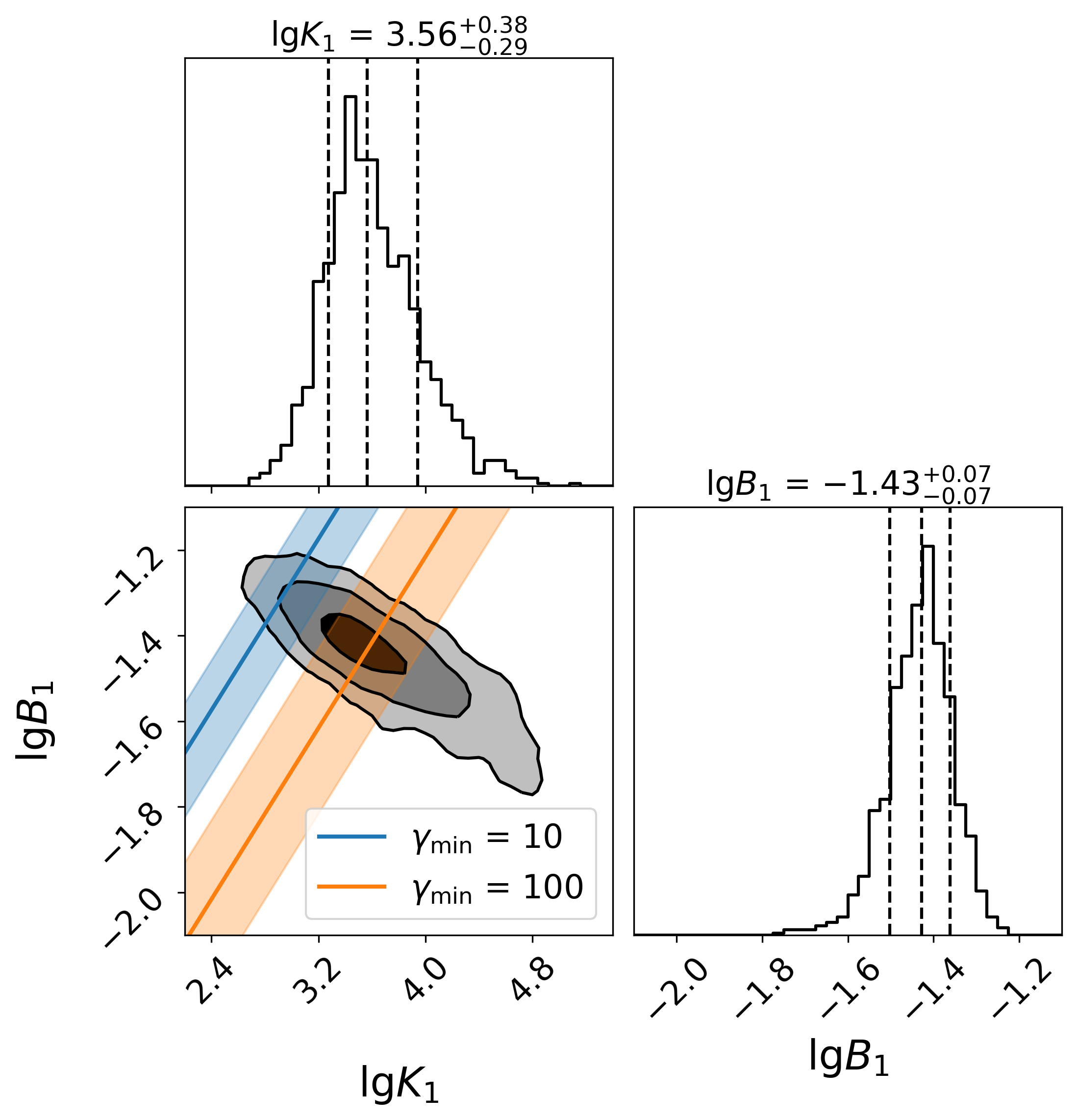}
      \caption{Distribution of the magnetic field and emitting particle density amplitude assuming jet speed in range 0.7--0.96c observed in pc-scale radio structure in \protect\cite{1999ApJ...519..108C}.
      The blue lines show equipartition for $\gamma_{\rm min}=10$ and orange lines -- for $\gamma_{\rm min}=100$. The lines show median and shaded area - uncertainty (95\%-credible interval) associated with uncertainty of the particles energy spectrum exponent $s=2\alpha+1$.}
       \label{fig:KB}
   \end{figure}

To infer the value of the fields we need an independent estimate of the Doppler factor as it is not constrained in our model. Velocity $\beta$ could be estimated to lie in a range 0.7--0.96 of component speeds observed in pc-scale radio structure \citep{1999ApJ...519..108C}.
Together with estimate of the jet viewing angle from \cite{2005MNRAS.363.1223C} we used the obtained posterior distribution of model parameters (Figure~\ref{fig:corner}) to calculate magnetic field and particle density at $r=1$ pc (Figure~\ref{fig:KB}). Here the lines correspond to the equipartition and are shown for $\gamma_{\rm min}$ = 10 and 100. When using relation between the jet opening angle and Lorentz factor derived in \cite{2013A&A...558A.144C}, the median and 95\% credible intervals are -1.41 and (-1.53, -1.29) for $\lg{B_1 {\rm [G]}}$ and 3.50 and (3.0, 4.18) for $\lg{K_1[{\rm cm^{-3}]}}$.

It is interesting to compare this estimate of the magnetic field with the traditional estimate that assumes equipartition \citep[e.g.][]{1998A&A...330...79L,2005ApJ...619...73H,2009MNRAS.400...26O}. \cite{2012A&A...545A.113P} obtained the following values of the core shift projected on the jet median position angle: 0.179 mas for 15.4--8.1 GHz\footnote{The posterior distribution of the model core shift (see Section~\ref{sec:cjcomponent} and Appendix~\ref{sec:extendingBK}) has median 0.165 mas and 95\% credible interval (0.159, 0.171)}, 0.064 mas for 15.4--8.4 GHz and 0.052 mas for 15.4--12.1 GHz. To calculate the magnetic field value one would assume equipartition \citep[thus $k_r = 1$,][]{1998A&A...330...79L} and $\alpha = 0.5$. With core shift value for 15.4--8.4 GHz that is consistent with one at higher frequency, we use equation (5) from \citep{2009MNRAS.400...26O} to obtain $\lg(B_1^{\rm eq}[{\rm G}])=$ -1.63 and -1.43 for $\beta=$ 0.7 and 0.96. For value of the core shift between 15.4 and 8.1 GHz the magnetic field is (-1.25, -1.05) for corresponding speeds. As discussed in \cite{2009MNRAS.400...26O} using any $\alpha$ other than 0.5 in derivation of \citep{2005ApJ...619...73H} results in $B_1^{\rm eq}$ dependence on $\gamma_{\rm min}$. Accounting for the estimated spectral index lowers $B_1^{\rm eq}$ with a factor $\approx$2 for values of $\gamma_{\rm min}$ from 10 to 100.

\subsection{Plasma and jet magnetization}
\label{sec:magnetization}

\begin{figure}
   \centering
   \includegraphics[width=\columnwidth, trim=0.3cm 0.5cm 0.3cm 0.3cm]{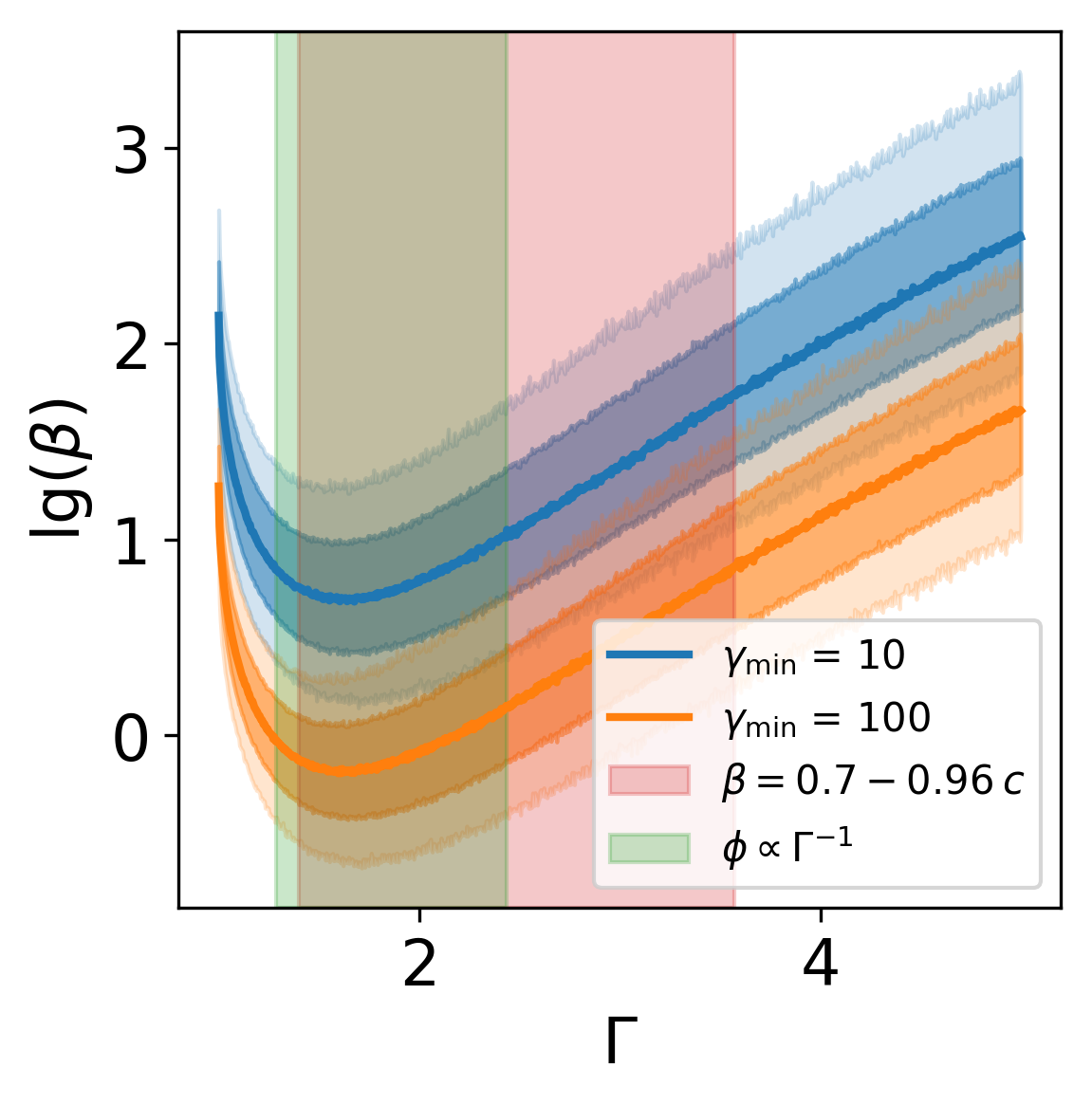}
      \caption{The dependence of the plasma $\beta_{\rm p}$ parameter (ratio of emitting particle energy density to that of magnetic field) on the jet Lorenz factor $\Gamma$. The blue color shows relation assuming $\gamma_{\rm min}=10$ and orange lines -- for $\gamma_{\rm min}=100$. The red stripe indicates range of the jet speeds 0.7--0.96c observed in pc-scale radio structure in \protect\cite{1999ApJ...519..108C}. Green lines show values of $\Gamma$ obtained from our posterior of $\phi_{\rm app}$, prior distribution of $\theta$ from \protect\citep{2005MNRAS.363.1223C} and relation between jet opening angle and bulk motion Lorenz factor obtained in \protect\citep{2013A&A...558A.144C}.}
       \label{fig:UbUeRatiovsGamma}
   \end{figure}

The dependence of the energy densities ratio of magnetic field and emitting particles on the jet Lorenz factor $\Gamma$ is presented in Figure~\ref{fig:UbUeRatiovsGamma}. The vertical red stripe represents the spread of the component speeds observed within VLBI structure by \cite{1999ApJ...519..108C}. With the sampled posterior distribution of our model parameters (Figure~\ref{fig:corner}) we also show the values of $\Gamma$ obtained using relation between jet opening angle and bulk motion Lorentz factor in \cite{2013A&A...558A.144C} (shown as vertical green stripe around $\Gamma\approx1.9$).

Our fit reveals that emitting plasma has $\beta_{\rm p} \approx 1$ for low energy cutoff $\gamma_{\rm min} = 100$, where the plasma parameter $\beta_{\rm p}$ is defined through the energy densities \citep{10.1093/mnras/stu1835}:
\begin{multline}
\label{eq:plasmabeta}
    \beta_{\rm p} = \frac{u_e}{u_B}\\
    u_e = n_{\rm pl}m_{\rm e}c^2\langle \gamma-1 \rangle (1+k_{\rm u})\\
    u_B = \frac{B^2}{8\pi}\\
\end{multline}
where $u_{\rm e}$ and $u_B$ --- emitting particle and magnetic field energy densities, $n_{\rm pl}$ --- number density of electrons in a power-law at a distance $r_1=1$ pc from the apex, $B$ --- magnetic field at this point (both in a plasma frame), $c$ --- speed of light, $k_{\rm u}$=0 --- contribution from particles outside of the power-law and ions. For $\gamma_{\rm min} = 10$ we obtain $\beta_{\rm p} \approx 10$ with 2$\sigma$ uncertainty 0.5 dex. If the break in particles energy spectrum exists $\gamma_{\rm br} \le 100$, $\beta_{\rm p}$ becomes lower and consistent with the equipartition. 

\begin{figure}
   \centering
   \includegraphics[width=\columnwidth, trim=0.3cm 0.5cm 0.3cm 0.3cm]{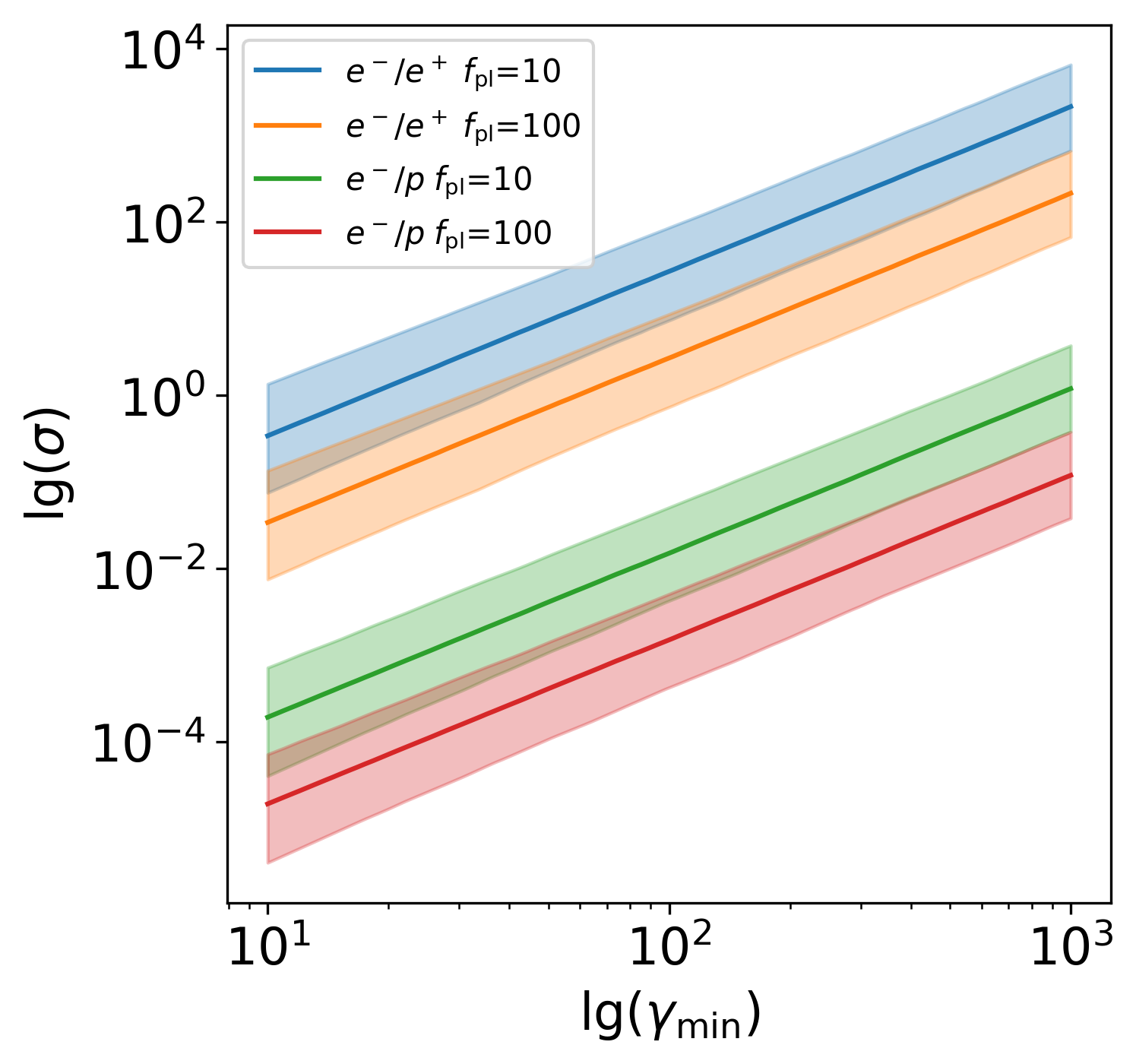}
      \caption{Dependence of the jet magnetization $\sigma$ on the lower cut off if the power law distribution of the emitting particles $\gamma_{\rm min}$ for $e^-/e^+$ and $e^-/p$ jet and fraction of particles outside of the power law $f_{\rm pl}$. The lines show median and shaded area - 95\% credible interval.}
       \label{fig:sigma}
   \end{figure}

Although we obtained $\beta_{\rm p} \approx 1$ for $\gamma_{\rm min} = 100$ it does not imply that jet magnetization parameter $\sigma \approx 1$, where $\sigma$ is the ratio of the Poynting flux to the kinetic energy flux in the black hole frame \citep{2005MNRAS.360..869L,2017MNRAS.468.2372N}:
\begin{equation}
    \sigma = \frac{B^2}{8\pi(n_{\rm e}m_{\rm e}+n_{\rm p}m_{\rm p})c^2}
    \label{eq:magnetization}
\end{equation}
where $n_{\rm p}$ is the number density (in jet frame) of protons at $r_1$, ($n_{\rm p} = 0$ for pairs jet), $m_{\rm p}$ - proton mass. We also assume that number density of electrons in a power-law $n_{\rm pl}$ is some fraction ($1/f_{\rm pl}$) of all electrons $n_{\rm e}$ and in case of normal plasma jet $n_{\rm e} = n_{\rm p}$. Figure~\ref{fig:sigma} shows the dependence of the jet magnetization $\sigma$ on the lower cutoff in the power law distribution of the emitting particles $\gamma_{\rm min}$ for $e^-/e^+$-- and $e^-/p$--jet. It is expected that acceleration and collimation of initially highly magnetized ($\sigma \gg 1$) jet takes place till the equipartition $\sigma \propto 1$ \citep[][and references therein]{2013MNRAS.429.1189P}.
Our results are consistent with this only for the pairs plasma jet and $\gamma_{\rm min} \le 100$. Normal plasma jet is particle-dominated up to $\gamma_{\rm min} \propto 1000$. 

Soft spectral index obtained in Section~\ref{sec:spixacc} could imply the magnetic reconnection as a particle acceleration process \citep{2014ApJ...783L..21S}. Magnetic reconnection in a both normal and pairs plasma results in various electrons spectra -- from $s\approx1.0$ to to $s\approx4$ depending on magnetization $\sigma$ and value of the guide magnetic field \citep{2014ApJ...783L..21S,Guo_2015,Guo_2016,2017ApJ...843L..27W,2018MNRAS.473.4840W}. It however requires $\sigma \gg 1$ to be efficient \citep{2018MNRAS.473.4840W}. 
Obtained values (Figure~\ref{fig:sigma}) are thus consistent with magnetic reconnection only for $e^-/e^+$ jet and $\gamma_{\rm min} \ge 100$.

\subsection{Jet composition}
\label{sec:composition}

\begin{figure}
   \centering
   \includegraphics[width=\columnwidth, trim=0.3cm 0.5cm 0.3cm 0.3cm]{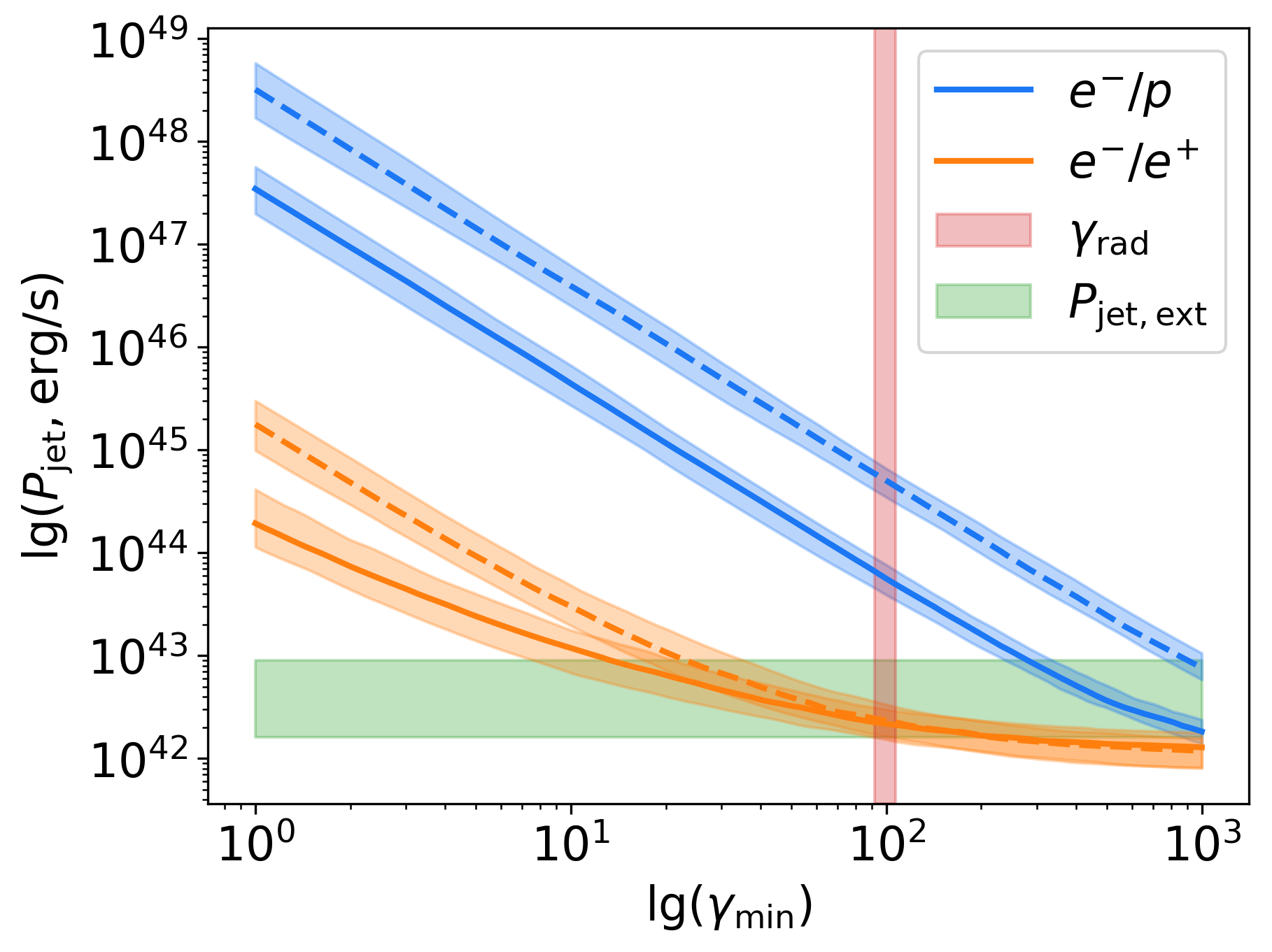}
      \caption{Dependence of the jet power for $e^-/p$ and $e^-/e^+$ plasma jet on the $\gamma_{\rm min}$. Vertical red stripe corresponds to constrain on $\gamma_{\rm min}$ from electrons emitting at SSA maximum at 15.4 GHz. Horizontal shaded area corresponds to constrain on jet power obtained in \citep{2010ApJ...720.1066C}. Continuous lines represent $f_{\rm pl}=10$ and dashed lines represent $f_{\rm pl}=100$. Lines show median of the posterior and shaded regions around the lines - corresponding 1$\sigma$ interval}.
       \label{fig:composition}
   \end{figure}

To constrain the jet composition we plot the dependence of the jet power $P_{\rm jet}$ estimated from our fit on $\gamma_{\rm min}$ for both normal and pairs plasma jet in Figure~\ref{fig:composition}. Here, following \cite{10.1093/mnras/stu1835} the jet power is presented as sum of components -- power in electrons/positrons, magnetic field and bulk-motion kinetic power:
\begin{multline}
        P_{\rm jet} = P_{\rm e} + P_{\rm B} + P_{\rm p} = \\
    = \pi c \beta (\Gamma r)^2 (\eta u_{\rm e}+\eta_{B}u_{B}) + \pi c \beta \Gamma (\Gamma-1) f_{\rm pl} n_{\rm pl} (m_{\rm p} + m_{\rm e})c^3 r^2
\end{multline}
where $\eta_B$ - magnetic adiabatic index (4/3 for tangled field), $\eta$ - average adiabatic index (4/3 < $\eta$ < 5/3). Lines represent median of posterior distribution and shaded regions -- corresponding $1\sigma$-interval. Continuous lines show $f_{\rm pl} = 10$ and dashed -- $f_{\rm pl} = 100$.

\cite{1994ApJ...422..542B} estimated the jet power of NGC~315 using extended radio structure from $4 \cdot 10^{42}$ up to $10^{43}$ erg/s assuming an equal contribution of the lobe internal energy and the work done on the lobe expansion. \cite{2006ApJ...648..200D} shown that the ratio of the work done on the lobe expansion to the lobe internal energy could be higher (with typical ratio values equal 3--10).
\cite{2010ApJ...720.1066C} estimated $\lg{P_{\rm jet}} = 6.58(+2.48-4.96)\cdot10^{42}$ erg/s using cavities in a X-ray emitting gas. As the radio lobes of NGC~315 are poorly confined, they assumed that volume of a cavity equals the volume of the corresponding radio lobe and extrapolated pressure profiles beyond the region observed in X-rays.
\cite{2009A&A...505..559M} used correlation between core radio luminosity and jet power from \citep{Heinz_2007} and obtained $P_{\rm jet} = 10^{44.15}$ erg/s with uncertainty 0.5 dex. However, this estimate was obtained for a sample of sources that are biased against beaming \citep{Heinz_2007} and probably is the upper limit for NGC~315.

Using expression for Lorentz factor $\gamma_{\rm rad}$ of particles emitting at synchrotron-self absorption peak $\nu$ ($\gamma_{\rm rad} \approx \sqrt{\nu/\nu_B}$, where the observed frequency $\nu$ is corrected for the Doppler factor and $\nu_B$ is Larmor frequency for given magnetic field) we can constrain electrons low-energy cutoff as $\gamma_{\rm min} < \gamma_{\rm rad}$. For flat spectrum sources $\nu/\nu_B$ is nearly constant and emission is dominated by electrons with the same $\gamma_{\rm rad}$ at all radii \citep{2002A&A...388.1106B,2019arXiv190203860B}.
In our model $\gamma_{\rm rad}$ = 99 for 15.4 GHz core. Thus $\gamma_{\rm min}$ can not be larger than $\approx$99.
This corresponds to the vertical red stripe in Figure~\ref{fig:composition} that shows 1$\sigma$ interval obtained from the posterior (Figure~\ref{fig:corner}).

\citep{1993MNRAS.264..228C,1996MNRAS.283..873R} found that if for $e^-/p$--jet lower energy cutoff $\gamma_{\rm min} < 100$ than such jet carries more energy than it seems to be dissipated. For pairs plasma jet $\gamma_{\rm min}$ should be $\approx 1$ for jets to carry the amount of the energy that is dissipated.
From Figure~\ref{fig:composition} it is apparent that electron-proton jet fails to explain both the necessary energetic and the lower energy cutoff in a power-law. Note, that possible break in a power-law at $\gamma_{\rm br}$ does not influence the jet power estimate for those $\gamma_{\rm min}$ that are close to the break. In case we are observing the steepen part of the synchrotron spectrum the break $\gamma_{\rm br}$ can not be larger than $\gamma_{\rm rad}\approx100$ for core region at 15.4 GHz that is well described by the non-uniform model with optically thin steep spectral index $\alpha \approx1.0$. However $\gamma_{\rm min}$ should not be much less or jet will carry the excessive power. In other words the possible break does not influence the conclusion concerning excess energetic of a normal plasma jet.

If estimate of $P_{\rm jet}$ from \cite{2010ApJ...720.1066C} is underestimated, i.e. value from \cite{2009A&A...505..559M} is applicable, possibly with the corresponding correction for beaming \citep[see also][for discussion of different methods of $P_{\rm jet}$ estimation]{10.1093/mnras/stw2960,2017ApJ...840...46I} than electron-proton jet can be reconciled with this jet power only if $f_{\rm pl}$ is not larger than $\propto$10. This implies quite efficient particles acceleration that is higher than expected for weakly magnetized relativistic shocks \citep[$\propto1\%$, ][]{2013ApJ...771...54S}. 

Using X-ray observation of NGC~315 we can check self-consistency of our model. As the number of particles depends on $s$ and $\gamma_{\rm min}$, with high $s$ and low $\gamma_{\rm min}$ we can get too many particles, which will produce excessive X-rays via Synchrotron-Self Compton (SSC) mechanism \citep{1996MNRAS.283..873R}. For particles density and size of the emitting region corresponding to the 15.4 GHz core we obtain 1 keV X-ray flux $1.2\cdot10^{-7}$ Jy assuming high-energy cutoff in  the synchrotron spectrum $\nu_{\rm b} = \exp(10)\cdot\nu_{\rm 15.4GHz}$ \citep{1992MNRAS.258..776G,1996MNRAS.283..873R}. \cite{2007MNRAS.380....2W} used \textit{Chandra} X-ray observations and obtained flux of the power law component of the central core region $(1.2\pm0.2) \cdot 10^{-7}$ Jy which they attributed to a jet.

\section{Components besides inhomogeneous model}
\label{sec:countercore}

\subsection{Counter-jet component}
\label{sec:cjcomponent}

As we noted in Section~\ref{sec:probmodel} the analysis of the residuals between the data and best fitted inhomogeneous model revealed the presence of the component in a counter-jet.
Parameters of this component are presented in Table~\ref{tab:countercore}. Mean values of the posterior distributions are presented as well as corresponding 1$\sigma$ credible intervals. 

\begin{figure}
   \centering
   \includegraphics[width=\columnwidth, trim=0.3cm 0.5cm 0.3cm 0.3cm]{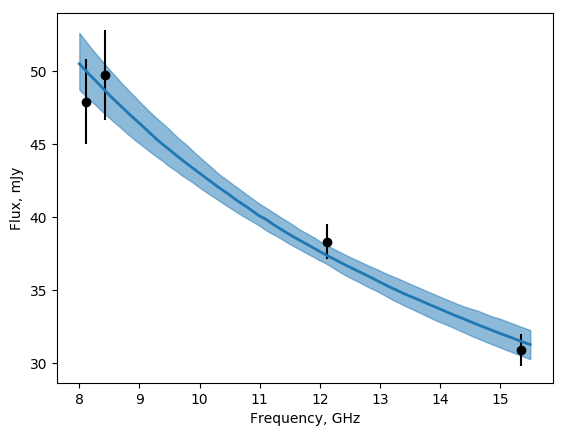}
   \caption{Spectrum of the component on the opposite from the visible jet side. The black dots and error bars show observed values and associated uncertainties. The blue line and band show prediction of the optically thin synchrotron model and corresponding 68\% confidence band.}
   \label{fig:cjspectra}
\end{figure}

The spectrum of the counter-jet component and its fit by an optically thin synchrotron model are presented in Figure~\ref{fig:cjspectra}. Optically thin spectral index is $0.73\pm0.09$ and agrees within 2$\sigma$ with the spectral index of our jet model. Note that for a counter-jet the same observed frequencies correspond to intrinsic frequencies $D^2$ times ($\approx$3--4 times) higher than intrinsic frequencies for the approaching jet.
This component can hardly be explained by a local brightening in a counter jet as the approaching jet is completely featureless.
Thus it should be the position of a $\tau\approx1$ surface, where optical depth $\tau$ is determined by intrinsic and(or) external absorption.

The position of this component relative to the cone apex of the jet changes little with frequency (see Table~\ref{tab:countercore}) and is consistent with being constant. This is shown in Figure~\ref{fig:expectedobservedtruecore} together with the same dependence for VLBI-core of the approaching jet according to our model. The intensity peak of the model emission was used as the VLBI core. As shown in Appendix~\ref{sec:extendingBK} to find the core position one has to solve an exponential equation for parameters $n$, $m$ and $\alpha$, which gives $\tau \approx3.1$ for maximum brightness. Then using our re-parametrization the corresponding dependence of the position of true core at some frequency $\nu_{\rm obs}$ is calculated as follows:
\begin{equation}
\label{eq:corepositio}
r_{\rm obs}(\nu_{\rm obs}) =  \left( \frac{\tau}{C_{\tau}(\alpha) A_2 \phi_{\rm app}}\right)^{-\frac{1}{n+m(1.5+\alpha)-1}} \nu_{\rm obs}^{-\frac{1}{k_r}}
\end{equation}
where $k_r = (n+m(1.5+\alpha)-1)/(2.5+\alpha)$. From the posterior (Figure~\ref{fig:corner}) we obtain the median $k_r=0.89$ and 95$\%$ credible interval for -- $(0.86, 0.91)$.

\begin{figure}
   \centering
   \includegraphics[width=\columnwidth, trim=0.3cm 0.5cm 0.3cm 0.3cm]{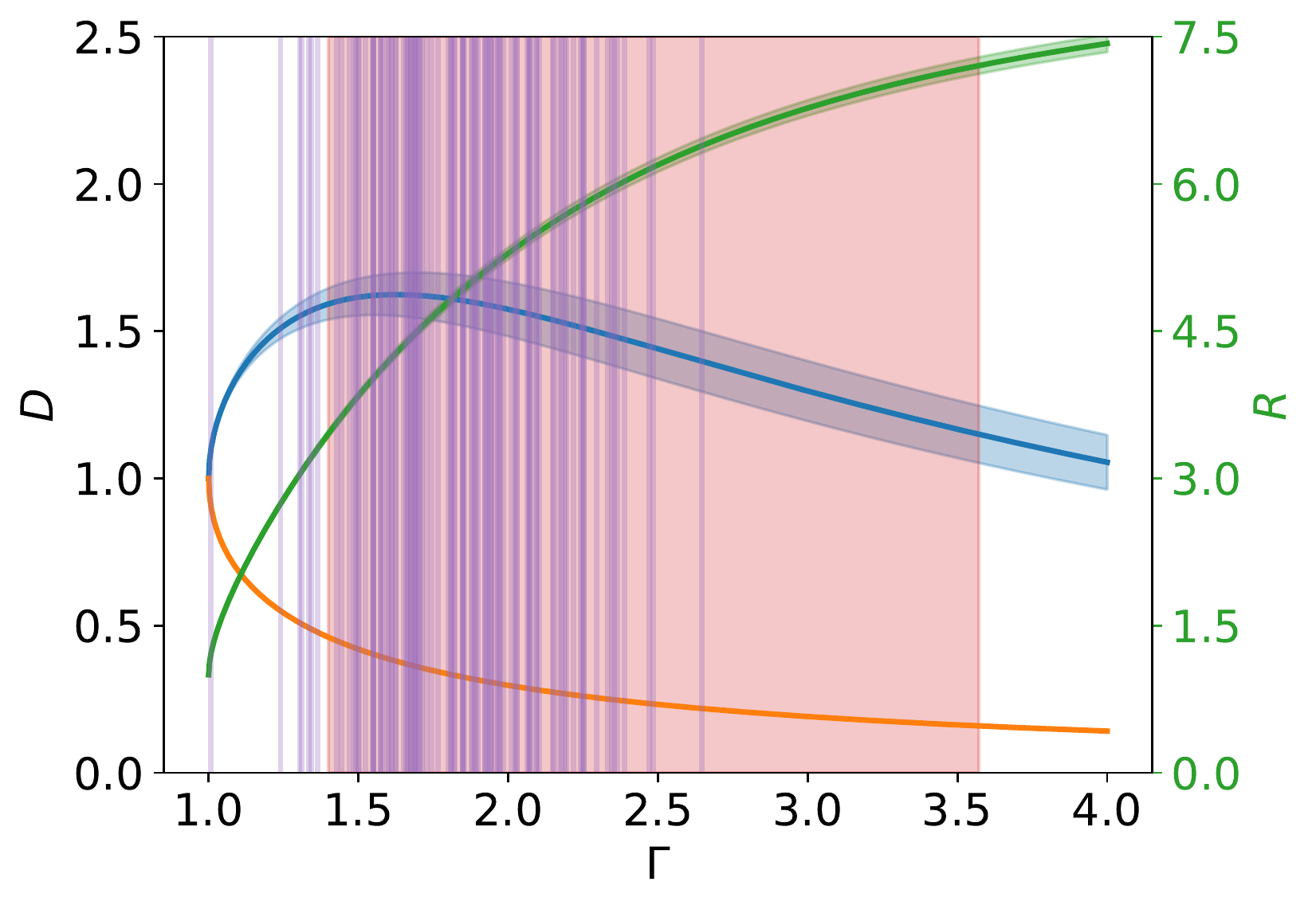}
   \caption{Dependence of the approaching jet Doppler factor (blue line), counter-jet Doppler factor (orange line) and their ratio $R$ (green line and right axis) on the jet Lorenz factor $\Gamma$. The vertical red stripe indicates range of the jet speeds 0.7--0.96c observed in pc-scale radio structure in \protect\cite{1999ApJ...519..108C}. The vertical purple lines show values of $\Gamma$ obtained from our posterior of $\phi_{\rm app}$, prior distribution of $\theta$ from \protect\citep{2005MNRAS.363.1223C} and relation between jet opening angle and bulk motion Lorenz factor obtained in \protect\citep{2013A&A...558A.144C}.}
   \label{fig:deltavsgamma}
\end{figure}

We also plot the expected position of the counter-core for a range of jet to counter-jet Doppler factor ratio. Here we assumed that synchrotron self-absorption is the only source of opacity. In Figure~\ref{fig:deltavsgamma} green line shows the dependence of jet to counter-jet Doppler factor ratio on the jet Lorenz factor $\Gamma$ for viewing angle $\theta = 38\pm2 \deg$ obtained in \citep{2005MNRAS.363.1223C}. The red wide vertical stripe indicates range of the jet speeds 0.7--0.96 observed in pc-scale radio structure in \cite{1999ApJ...519..108C}. Purple vertical lines show values of $\Gamma$ obtained from the joint posterior of $\phi_{\rm app}$, $\theta$ from \citep{2005MNRAS.363.1223C} and relation between the jet opening angle $\phi$ and bulk motion Lorenz factor $\Gamma$ obtained in \citep{2013A&A...558A.144C}.

\begin{table}
\begin{center}
\caption{Parameters of the counter-jet component.}
\label{tab:countercore}
\begin{tabular}{c@{~~~}c@{~~~}c@{~~~}c}
\hline
Frequency & Flux density & $r$  \\
GHz     & mJy  & mas \\
\hline
 8.1  & 47.9$\pm$2.9 & -0.173$\pm$0.015  \\
 8.4  & 49.7$\pm$3.1 & -0.168$\pm$0.013  \\
 12.1  & 38.3$\pm$1.2 & -0.220$\pm$0.008 \\
 15.4  & 30.9$\pm$1.1 & -0.185$\pm$0.007  \\

\hline
\end{tabular}
\end{center}
{\bf Column designation:}
Col.~1~-- Frequency, GHz,
Col.~2~-- Flux density, mJy,
Col.~3~-- Distance from cone apex, mas.
\end{table}

As shown in \citep{1998A&A...330...79L}\footnote{In Section~2.4 of that paper $r_{\rm core} \approx \nu^{-1/2.5}$ should be instead of $\nu^{-2.5}$. See also \citep{2004A&A...426..481K}.} in the environment with a steep density gradients $k_r$ as high as 2.5 are possible. In the presence of strong external density gradients values of $k_r$ are generally even higher. In our case position of the counter-jet component is nearly constant within the errors and requires $k_{r} \gg 1$. We thus attribute the counter-jet component to ``core'' for which the dominating opacity is due to the external absorber with extremely high ionized gas density gradient and size $\approx0.1$ pc. This is close to the size of the Broad Line Region (BLR). \cite{1999ApJ...525..673B} detected broad polarized H$\alpha$ and H$\beta$ lines in optical spectrum of NGC~315 suggesting an obscured BLR to exists in its nucleus.

The core of counter-jet and its frequency dependent shift were detected before in e.g. \cite{2013EPJWC..6108004H} in radio galaxy NGC~4261 using multi-frequency (7 bands from 1.4 to 43 GHz) VLBA data. Authors concluded that neither pure synchrotron self-absorption (SSA) nor pure external (free-free) absorption (FFA) can explain the shift. \cite{2004A&A...426..481K} obtained for NGC~1051 the value of $k_r$ as high as 6.8 for counter jet and from 2.1 to 4.1 for the approaching jet. They attributed this to dominating contribution of FFA to the opacity on the counter-jet side. 
\cite{2013EPJWC..6108004H} found that lower and higher frequency position of the counter-jet core are explained by SSA and its position at medium frequency band needs contribution from the external absorption. They invoke model of outer disk and inner radiation inefficient accretion flow (RIAF) to explain such behaviour.
In our data with only 3 well separated relatively high-frequency bands the counter-jet core is even not apparent in the reconstructed CLEAN-images and in simplistic Gaussians models due to its weakness and closeness to the bright core of approaching jet. Thus it can be recovered only through the detailed modelling of the source structure.

\subsection{Constrain on plasma velocity}
\label{sec:vconstrain}
\begin{figure}
   \centering
   \includegraphics[width=\columnwidth, trim=0.3cm 0.5cm 0.3cm 0.3cm]{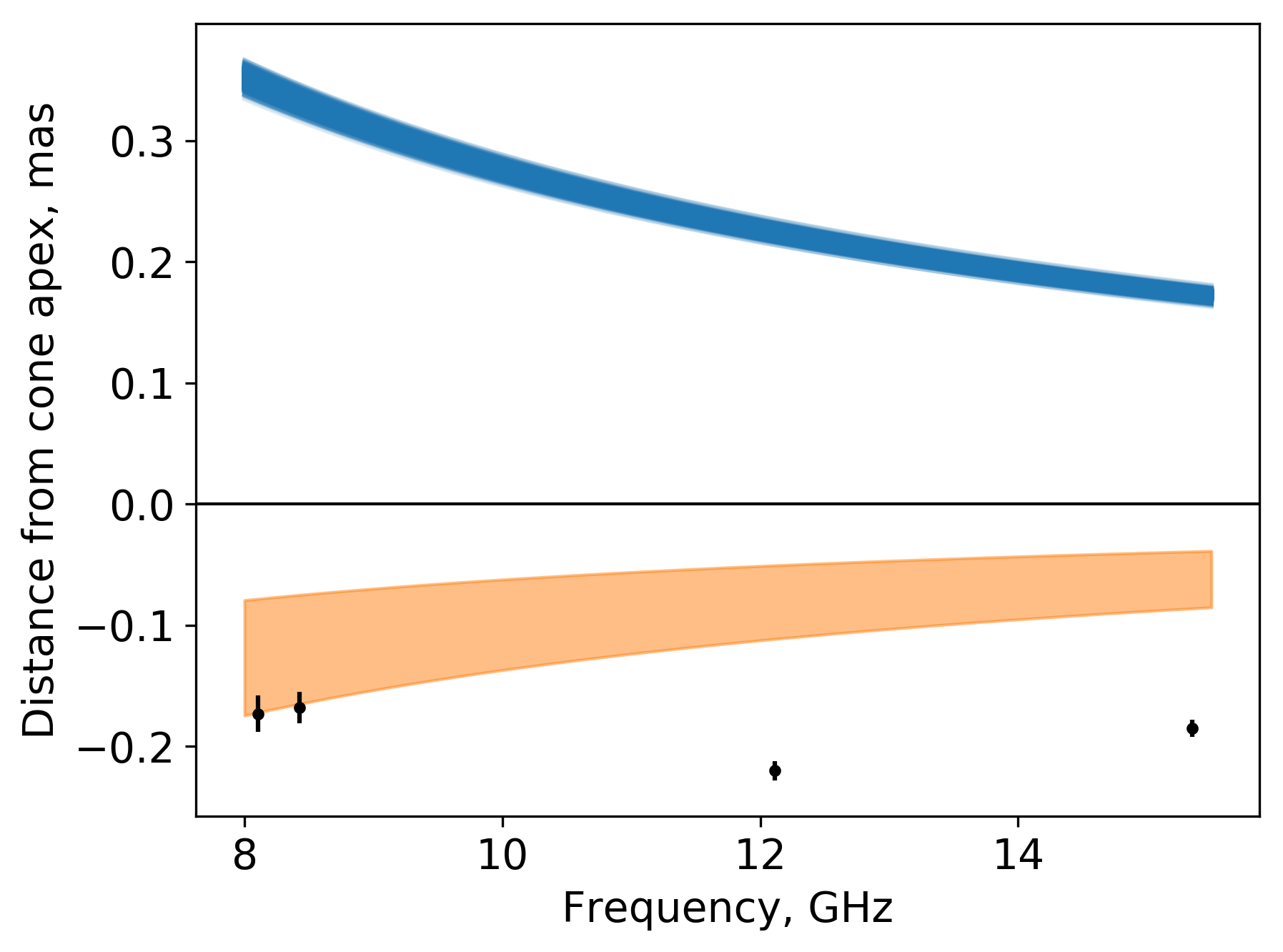}
   \caption{Thin blue lines show realizations of the core shift frequency dependence obtained from our fitted model. The orange stripe shows shift of the core in a counter-jet predicted from SSA model (Eq.~\ref{eq:corepositio}) with Doppler factors ratio $R$ in range 2.5--7.0. Black dots and error bars show position of the counter-jet component relative to cone apex (Column~3 in Table~\ref{tab:countercore}).}
   \label{fig:expectedobservedtruecore}
\end{figure}

We can constrain the ratio of the Doppler factors $R$ for approaching jet and counter-jet by the position of the counter-jet component at the lowest frequencies. Indeed, visible component can not be closer to jet apex than the region with optical depth $\tau \sim 1$. The depth is determined by the opacity -- both internal and external. In Figure~\ref{fig:expectedobservedtruecore} the orange stripe corresponds to $R \in (2.5, 7.0)$ or equivalently $\beta \in (0.7, 0.96)$ as estimated by \cite{2005MNRAS.363.1223C}. It turns out that independently from these estimates we can constrain $R > 2.5$, i.e. $\beta > 0.7$ at $r\approx$0.1 pc for viewing angle $\theta = 38\deg$. We stress that this constrains the plasma velocity, while VLBI kinematics could measure the pattern speed velocity \citep[e.g.][]{2014ApJ...787..151C}. 
Note that possible parabolic shape of the jet at its origin only strengthens the constrain. In this case the apex of the jet is closer to the counter-jet side than it follows from the conical geometry of our model (Kovalev+ in prep). Thus the observed counter-jet core is also closer to the jet apex. This implies that the ratio of the Doppler factors and corresponding lower limit on the plasma velocity should be even higher.

\subsection{Component in the approaching jet}
\label{sec:comp4mas}
The residuals between data and model imply the existence of a component at distance $\approx4$ mas from core at all observed frequency bands. The parameters of this component obtained with \texttt{Difmap} modelling of the residual visibilities are listed in Table~\ref{tab:jetcomp}. Component shows optically thin spectrum with $\alpha_{8.1}^{12.1}$ = $0.55\pm0.20$ and $\alpha_{12.1}^{15.4}$ = $4.0\pm0.8$. The first agrees with our model spectral index at $2\sigma$ level, but the second is significantly steeper. Velocity $\beta = 0.7$ corresponds to angular apparent speed $\mu_{\rm app}$ = 0.85 mas/year for the adopted viewing angle 38$^\circ$. Thus, the distance of 4 mas corresponds to travelling time from the core 4.7 years for $\beta = 0.7$ and 3.6 years for $\beta = 0.9$ assuming constant propagation speed. There are no signs of flaring activity near epoch 2001 in Figure~\ref{fig:lightcurve}, that is for 5 years preceding the observations we use. However the light curve cadence is too low to exclude flares with duration less than a year.

\begin{table}
\begin{center}
\caption{Parameters of the jet component.}
\label{tab:jetcomp}
\begin{tabular}{c@{~~~}c@{~~~}c@{~~~}c}
\hline
Frequency & Flux density & $r$  \\
GHz     & mJy  & mas \\
\hline
 8.1  & 11.6$\pm$0.6 & 4.35$\pm$0.05  \\
 8.4  & 9.5$\pm$0.6 & 4.39$\pm$0.05  \\
 12.1  & 9.3$\pm$0.6 & 4.49$\pm$0.04 \\
 15.4  & 3.6$\pm$0.6 & 4.83$\pm$0.06  \\

\hline
\end{tabular}
\end{center}
{\bf Column designation:}
Col.~1~-- Frequency, GHz,
Col.~2~-- Flux, mJy,
Col.~3~-- Distance from phase center, mas.
\end{table}

\section{Conclusions}
\label{sec:conslusion}

For the first time we fitted a physical jet model directly to the interferometric observables --- visibilities.
We used a non-uniform conical jet model which was generalized from \cite{1979ApJ...232...34B} by allowing arbitrary spectral index and exponents in power-law dependencies of the magnetic field and particle density on the distance. To break the degeneracies in model parameters we used multi-frequency VLBA data of a radio galaxy NGC~315 and assumed a constant magnetic field to particle energy density ratio. The observed data at all frequencies is well described by the model. We found that electron-positron jet is consistent with an independently derived jet power. The emitting plasma is consistent with the equipartition between magnetic field and emitting particles. The jet magnetization $\sigma$ is  0.1--1 depending on the fraction of pairs in a power-law.
Electron-proton jet could be reconciled with the data only if the jet power inferred from cavities in a X-ray emitting gas is underestimated for NGC~315 and particles acceleration is highly efficient.
Such jet is particle dominated with $\sigma \approx 10^{-3}$--$10^{-2}$.

We found a weak component on a counter-jet side in the residuals between the observed data and best model at all frequency bands. We attribute it to the external absorber with steep density gradient, possibly a border region of the BLR with a significantly ionized plasma. Position of the component at the 15.4 GHz constrains the plasma flow velocity $\beta > 0.7$ at distance $r\approx0.1$ pc from the central engine.

\section*{Acknowledgements}


We thank the anonymous referee for helpful comments and suggestions. We also thank Marina Butuzova and Mikhail Lisakov for reading the manuscript and helpful comments.
This work is supported by Russian Science Foundation grant 16-12-10481.
This research has made use of data from the MOJAVE database that is
maintained by the MOJAVE team \cite{mojave6kinematicsanalysis}.

This research has made use of NASA's Astrophysics Data System.

This research made use of \textit{Astropy}, a community-developed core Python package for Astronomy \citep{2013A&A...558A..33A}, \textit{Numpy} \citep{numpy}, \textit{Scipy} \citep{scipy}, \textit{numba} \citep{Lam:2015:NLP:2833157.2833162}, \textit{FINUFFT} \citep{2018arXiv180806736B}, \textit{emcee} \citep{2013PASP..125..306F}, \textit{PolyChord} \citep{polychord1,polychord2}. \textit{Matplotlib} Python package \citep{Hunter:2007} was used for generating all plots in this paper.




\bibliographystyle{mnras}
\bibliography{paper} 

\begin{thebibliography}{}
\makeatletter
\relax
\def\mn@urlcharsother{\let\do\@makeother \do\$\do\&\do\#\do\^\do\_\do\%\do\~}
\def\mn@doi{\begingroup\mn@urlcharsother \@ifnextchar [ {\mn@doi@}
  {\mn@doi@[]}}
\def\mn@doi@[#1]#2{\def\@tempa{#1}\ifx\@tempa\@empty \href
  {http://dx.doi.org/#2} {doi:#2}\else \href {http://dx.doi.org/#2} {#1}\fi
  \endgroup}
\def\mn@eprint#1#2{\mn@eprint@#1:#2::\@nil}
\def\mn@eprint@arXiv#1{\href {http://arxiv.org/abs/#1} {{\tt arXiv:#1}}}
\def\mn@eprint@dblp#1{\href {http://dblp.uni-trier.de/rec/bibtex/#1.xml}
  {dblp:#1}}
\def\mn@eprint@#1:#2:#3:#4\@nil{\def\@tempa {#1}\def\@tempb {#2}\def\@tempc
  {#3}\ifx \@tempc \@empty \let \@tempc \@tempb \let \@tempb \@tempa \fi \ifx
  \@tempb \@empty \def\@tempb {arXiv}\fi \@ifundefined
  {mn@eprint@\@tempb}{\@tempb:\@tempc}{\expandafter \expandafter \csname
  mn@eprint@\@tempb\endcsname \expandafter{\@tempc}}}

\bibitem[\protect\citeauthoryear{{Andrae}, {Schulze-Hartung}  \&
  {Melchior}}{{Andrae} et~al.}{2010}]{2010arXiv1012.3754A}
{Andrae} R.,  {Schulze-Hartung} T.,   {Melchior} P.,  2010, arXiv e-prints,
  \href {https://ui.adsabs.harvard.edu/abs/2010arXiv1012.3754A} {p.
  arXiv:1012.3754}

\bibitem[\protect\citeauthoryear{{Astropy Collaboration} et~al.,}{{Astropy
  Collaboration} et~al.}{2013}]{2013A&A...558A..33A}
{Astropy Collaboration} et~al., 2013, \mn@doi [\aap]
  {10.1051/0004-6361/201322068}, \href
  {http://adsabs.harvard.edu/abs/2013A%26A...558A..33A} {558, A33}

\bibitem[\protect\citeauthoryear{{Barnett}, {Magland}  \&
  {Klinteberg}}{{Barnett} et~al.}{2018}]{2018arXiv180806736B}
{Barnett} A.~H.,  {Magland} J.~F.,   {Klinteberg} L.~a.,  2018, arXiv e-prints,
  \href {https://ui.adsabs.harvard.edu/abs/2018arXiv180806736B} {p.
  arXiv:1808.06736}

\bibitem[\protect\citeauthoryear{{Barth}, {Filippenko}  \& {Moran}}{{Barth}
  et~al.}{1999}]{1999ApJ...525..673B}
{Barth} A.~J.,  {Filippenko} A.~V.,   {Moran} E.~C.,  1999, \mn@doi [\apj]
  {10.1086/307941}, \href {http://adsabs.harvard.edu/abs/1999ApJ...525..673B}
  {525, 673}

\bibitem[\protect\citeauthoryear{{Bayes} \& {Price}}{{Bayes} \&
  {Price}}{1763}]{bayes}
{Bayes} T.,  {Price} R.,  1763, Philosophical Transactions of the Royal Society
  of London, 53, 370

\bibitem[\protect\citeauthoryear{{Beckert} \& {Falcke}}{{Beckert} \&
  {Falcke}}{2002}]{2002A&A...388.1106B}
{Beckert} T.,  {Falcke} H.,  2002, \mn@doi [\aap] {10.1051/0004-6361:20020484},
  \href {http://adsabs.harvard.edu/abs/2002A%26A...388.1106B} {388, 1106}

\bibitem[\protect\citeauthoryear{{Bicknell}}{{Bicknell}}{1994}]{1994ApJ...422..542B}
{Bicknell} G.~V.,  1994, \mn@doi [\apj] {10.1086/173748}, \href
  {http://adsabs.harvard.edu/abs/1994ApJ...422..542B} {422, 542}

\bibitem[\protect\citeauthoryear{{Bjornsson}}{{Bjornsson}}{2019}]{2019arXiv190203860B}
{Bjornsson} C.-I.,  2019, arXiv e-prints, \href
  {https://ui.adsabs.harvard.edu/\#abs/2019arXiv190203860B} {p.
  arXiv:1902.03860}

\bibitem[\protect\citeauthoryear{{Blandford} \& {K{\"o}nigl}}{{Blandford} \&
  {K{\"o}nigl}}{1979}]{1979ApJ...232...34B}
{Blandford} R.~D.,  {K{\"o}nigl} A.,  1979, \mn@doi [\apj] {10.1086/157262},
  \href {http://adsabs.harvard.edu/abs/1979ApJ...232...34B} {232, 34}

\bibitem[\protect\citeauthoryear{{Bridle}, {Davis}, {Meloy}, {Fomalont},
  {Strom}  \& {Willis}}{{Bridle} et~al.}{1976}]{1976Natur.262..179B}
{Bridle} A.~H.,  {Davis} M.~M.,  {Meloy} D.~A.,  {Fomalont} E.~B.,  {Strom}
  R.~G.,   {Willis} A.~G.,  1976, \mn@doi [\nat] {10.1038/262179a0}, \href
  {http://adsabs.harvard.edu/abs/1976Natur.262..179B} {262, 179}

\bibitem[\protect\citeauthoryear{Briggs}{Briggs}{1995}]{briggs}
Briggs D.~S.,  1995, PhD thesis, The New Mexico Institue of Mining and
  Technology, Socorro, New Mexico

\bibitem[\protect\citeauthoryear{{Canvin}, {Laing}, {Bridle}  \&
  {Cotton}}{{Canvin} et~al.}{2005}]{2005MNRAS.363.1223C}
{Canvin} J.~R.,  {Laing} R.~A.,  {Bridle} A.~H.,   {Cotton} W.~D.,  2005,
  \mn@doi [\mnras] {10.1111/j.1365-2966.2005.09537.x}, \href
  {http://adsabs.harvard.edu/abs/2005MNRAS.363.1223C} {363, 1223}

\bibitem[\protect\citeauthoryear{{Cavagnolo}, {McNamara}, {Nulsen}, {Carilli},
  {Jones}  \& {B{\^i}rzan}}{{Cavagnolo} et~al.}{2010}]{2010ApJ...720.1066C}
{Cavagnolo} K.~W.,  {McNamara} B.~R.,  {Nulsen} P.~E.~J.,  {Carilli} C.~L.,
  {Jones} C.,   {B{\^i}rzan} L.,  2010, \mn@doi [\apj]
  {10.1088/0004-637X/720/2/1066}, \href
  {http://adsabs.harvard.edu/abs/2010ApJ...720.1066C} {720, 1066}

\bibitem[\protect\citeauthoryear{{Celotti} \& {Fabian}}{{Celotti} \&
  {Fabian}}{1993}]{1993MNRAS.264..228C}
{Celotti} A.,  {Fabian} A.~C.,  1993, \mn@doi [\mnras]
  {10.1093/mnras/264.1.228}, \href
  {http://adsabs.harvard.edu/abs/1993MNRAS.264..228C} {264, 228}

\bibitem[\protect\citeauthoryear{{Clausen-Brown}, {Savolainen}, {Pushkarev},
  {Kovalev}  \& {Zensus}}{{Clausen-Brown} et~al.}{2013}]{2013A&A...558A.144C}
{Clausen-Brown} E.,  {Savolainen} T.,  {Pushkarev} A.~B.,  {Kovalev} Y.~Y.,
  {Zensus} J.~A.,  2013, \mn@doi [\aap] {10.1051/0004-6361/201322203}, \href
  {http://adsabs.harvard.edu/abs/2013A%26A...558A.144C} {558, A144}

\bibitem[\protect\citeauthoryear{{Cohen} et~al.,}{{Cohen}
  et~al.}{2014}]{2014ApJ...787..151C}
{Cohen} M.~H.,  et~al., 2014, \mn@doi [\apj] {10.1088/0004-637X/787/2/151},
  \href {http://adsabs.harvard.edu/abs/2014ApJ...787..151C} {787, 151}

\bibitem[\protect\citeauthoryear{{Cotton}, {Feretti}, {Giovannini}, {Lara}  \&
  {Venturi}}{{Cotton} et~al.}{1999}]{1999ApJ...519..108C}
{Cotton} W.~D.,  {Feretti} L.,  {Giovannini} G.,  {Lara} L.,   {Venturi} T.,
  1999, \mn@doi [\apj] {10.1086/307358}, \href
  {http://adsabs.harvard.edu/abs/1999ApJ...519..108C} {519, 108}

\bibitem[\protect\citeauthoryear{Davies}{Davies}{1995}]{davis1995}
Davies P.,  1995, \mn@doi [Statistica Neerlandica]
  {10.1111/j.1467-9574.1995.tb01464.x}, 49, 185

\bibitem[\protect\citeauthoryear{{De Young}}{{De
  Young}}{2006}]{2006ApJ...648..200D}
{De Young} D.~S.,  2006, \mn@doi [\apj] {10.1086/505861}, \href
  {https://ui.adsabs.harvard.edu/\#abs/2006ApJ...648..200D} {648, 200}

\bibitem[\protect\citeauthoryear{{Ekers}, {Fanti}  \& {Miley}}{{Ekers}
  et~al.}{1983}]{1983A&A...120..297E}
{Ekers} R.~D.,  {Fanti} R.,   {Miley} G.~K.,  1983, \aap, \href
  {http://adsabs.harvard.edu/abs/1983A%26A...120..297E} {120, 297}

\bibitem[\protect\citeauthoryear{{Fanaroff} \& {Riley}}{{Fanaroff} \&
  {Riley}}{1974}]{1974MNRAS.167P..31F}
{Fanaroff} B.~L.,  {Riley} J.~M.,  1974, \mn@doi [\mnras]
  {10.1093/mnras/167.1.31P}, \href
  {http://adsabs.harvard.edu/abs/1974MNRAS.167P..31F} {167, 31P}

\bibitem[\protect\citeauthoryear{{Finke}}{{Finke}}{2019}]{finke}
{Finke} J.~D.,  2019, \mn@doi [\apj] {10.3847/1538-4357/aaf00c}, \href
  {https://ui.adsabs.harvard.edu/\#abs/2019ApJ...870...28F} {870, 28}

\bibitem[\protect\citeauthoryear{{Foreman-Mackey}, {Hogg}, {Lang}  \&
  {Goodman}}{{Foreman-Mackey} et~al.}{2013}]{2013PASP..125..306F}
{Foreman-Mackey} D.,  {Hogg} D.~W.,  {Lang} D.,   {Goodman} J.,  2013, \mn@doi
  [\pasp] {10.1086/670067}, \href
  {http://adsabs.harvard.edu/abs/2013PASP..125..306F} {125, 306}

\bibitem[\protect\citeauthoryear{{Fromm} et~al.,}{{Fromm}
  et~al.}{2019}]{2019arXiv190400106F}
{Fromm} C.~M.,  et~al., 2019, arXiv e-prints, \href
  {https://ui.adsabs.harvard.edu/\#abs/2019arXiv190400106F} {p.
  arXiv:1904.00106}

\bibitem[\protect\citeauthoryear{Gelman, Carlin, Stern, Dunson, Vehtari  \&
  Rubin}{Gelman et~al.}{2013}]{gelman2013bayesian}
Gelman A.,  Carlin J.,  Stern H.,  Dunson D.,  Vehtari A.,   Rubin D.,  2013,
  Bayesian Data Analysis, Third Edition.
Chapman \& Hall/CRC Texts in Statistical Science, Taylor \& Francis, \url
  {https://books.google.ru/books?id=ZXL6AQAAQBAJ}

\bibitem[\protect\citeauthoryear{{Ghisellini}, {Celotti}, {George}  \&
  {Fabian}}{{Ghisellini} et~al.}{1992}]{1992MNRAS.258..776G}
{Ghisellini} G.,  {Celotti} A.,  {George} I.~M.,   {Fabian} A.~C.,  1992,
  \mn@doi [\mnras] {10.1093/mnras/258.4.776}, \href
  {http://adsabs.harvard.edu/abs/1992MNRAS.258..776G} {258, 776}

\bibitem[\protect\citeauthoryear{Guo, Liu, Daughton  \& Li}{Guo
  et~al.}{2015}]{Guo_2015}
Guo F.,  Liu Y.-H.,  Daughton W.,   Li H.,  2015, \mn@doi [The Astrophysical
  Journal] {10.1088/0004-637x/806/2/167}, 806, 167

\bibitem[\protect\citeauthoryear{Guo et~al.,}{Guo et~al.}{2016}]{Guo_2016}
Guo F.,  et~al., 2016, \mn@doi [The Astrophysical Journal]
  {10.3847/2041-8205/818/1/l9}, 818, L9

\bibitem[\protect\citeauthoryear{{Haga}, {Doi}, {Murata}, {Sudou}, {Kameno},
  {Hada}  \& {Nagai}}{{Haga} et~al.}{2013}]{2013EPJWC..6108004H}
{Haga} T.,  {Doi} A.,  {Murata} Y.,  {Sudou} H.,  {Kameno} S.,  {Hada} K.,
  {Nagai} H.,  2013, in European Physical Journal Web of Conferences. p. 08004,
  \mn@doi{10.1051/epjconf/20136108004}

\bibitem[\protect\citeauthoryear{{Handley}, {Hobson}  \& {Lasenby}}{{Handley}
  et~al.}{2015a}]{polychord1}
{Handley} W.~J.,  {Hobson} M.~P.,   {Lasenby} A.~N.,  2015a, \mn@doi [\mnras]
  {10.1093/mnrasl/slv047}, \href
  {https://ui.adsabs.harvard.edu/\#abs/2015MNRAS.450L..61H} {450, L61}

\bibitem[\protect\citeauthoryear{{Handley}, {Hobson}  \& {Lasenby}}{{Handley}
  et~al.}{2015b}]{polychord2}
{Handley} W.~J.,  {Hobson} M.~P.,   {Lasenby} A.~N.,  2015b, \mn@doi [\mnras]
  {10.1093/mnras/stv1911}, \href
  {https://ui.adsabs.harvard.edu/\#abs/2015MNRAS.453.4384H} {453, 4384}

\bibitem[\protect\citeauthoryear{Heinz, Merloni  \& Schwab}{Heinz
  et~al.}{2007}]{Heinz_2007}
Heinz S.,  Merloni A.,   Schwab J.,  2007, \mn@doi [The Astrophysical Journal]
  {10.1086/513507}, 658, L9

\bibitem[\protect\citeauthoryear{{Hinshaw} et~al.,}{{Hinshaw}
  et~al.}{2013}]{2013ApJS..208...19H}
{Hinshaw} G.,  et~al., 2013, \mn@doi [\apjs] {10.1088/0067-0049/208/2/19},
  \href {http://adsabs.harvard.edu/abs/2013ApJS..208...19H} {208, 19}

\bibitem[\protect\citeauthoryear{{Hirotani}}{{Hirotani}}{2005}]{2005ApJ...619...73H}
{Hirotani} K.,  2005, \mn@doi [\apj] {10.1086/426497}, \href
  {http://adsabs.harvard.edu/abs/2005ApJ...619...73H} {619, 73}

\bibitem[\protect\citeauthoryear{{Ho}, {Filippenko}, {Sargent}  \& {Peng}}{{Ho}
  et~al.}{1997}]{1997ApJS..112..391H}
{Ho} L.~C.,  {Filippenko} A.~V.,  {Sargent} W.~L.~W.,   {Peng} C.~Y.,  1997,
  \mn@doi [\apjs] {10.1086/313042}, \href
  {http://adsabs.harvard.edu/abs/1997ApJS..112..391H} {112, 391}

\bibitem[\protect\citeauthoryear{{Hovatta} et~al.,}{{Hovatta}
  et~al.}{2014}]{2014AJ....147..143H}
{Hovatta} T.,  et~al., 2014, \mn@doi [\aj] {10.1088/0004-6256/147/6/143}, \href
  {http://adsabs.harvard.edu/abs/2014AJ....147..143H} {147, 143}

\bibitem[\protect\citeauthoryear{Hunter}{Hunter}{2007}]{Hunter:2007}
Hunter J.~D.,  2007, \mn@doi [Computing In Science \& Engineering]
  {10.1109/MCSE.2007.55}, 9, 90

\bibitem[\protect\citeauthoryear{{Inoue}, {Doi}, {Tanaka}, {Sikora}  \&
  {Madejski}}{{Inoue} et~al.}{2017}]{2017ApJ...840...46I}
{Inoue} Y.,  {Doi} A.,  {Tanaka} Y.~T.,  {Sikora} M.,   {Madejski} G.~M.,
  2017, \mn@doi [\apj] {10.3847/1538-4357/aa6b57}, \href
  {https://ui.adsabs.harvard.edu/abs/2017ApJ...840...46I} {840, 46}

\bibitem[\protect\citeauthoryear{Jones, Oliphant, Peterson  et~al.}{Jones
  et~al.}{2001}]{scipy}
Jones E.,  Oliphant T.,  Peterson P.,   et~al., 2001, {SciPy}: Open source
  scientific tools for {Python}, \url {http://www.scipy.org/}

\bibitem[\protect\citeauthoryear{{Kadler}, {Ros}, {Lobanov}, {Falcke}  \&
  {Zensus}}{{Kadler} et~al.}{2004}]{2004A&A...426..481K}
{Kadler} M.,  {Ros} E.,  {Lobanov} A.~P.,  {Falcke} H.,   {Zensus} J.~A.,
  2004, \mn@doi [\aap] {10.1051/0004-6361:20041051}, \href
  {https://ui.adsabs.harvard.edu/#abs/2004A&A...426..481K} {426, 481}

\bibitem[\protect\citeauthoryear{{Kardashev}}{{Kardashev}}{1962}]{1962SvA.....6..317K}
{Kardashev} N.~S.,  1962, \sovast, \href
  {http://adsabs.harvard.edu/abs/1962SvA.....6..317K} {6, 317}

\bibitem[\protect\citeauthoryear{{Konigl}}{{Konigl}}{1981}]{1981ApJ...243..700K}
{Konigl} A.,  1981, \mn@doi [\apj] {10.1086/158638}, \href
  {http://adsabs.harvard.edu/abs/1981ApJ...243..700K} {243, 700}

\bibitem[\protect\citeauthoryear{{Kovalev} et~al.,}{{Kovalev}
  et~al.}{2005}]{2005AJ....130.2473K}
{Kovalev} Y.~Y.,  et~al., 2005, \mn@doi [\aj] {10.1086/497430}, \href
  {http://adsabs.harvard.edu/abs/2005AJ....130.2473K} {130, 2473}

\bibitem[\protect\citeauthoryear{Lam, Pitrou  \& Seibert}{Lam
  et~al.}{2015}]{Lam:2015:NLP:2833157.2833162}
Lam S.~K.,  Pitrou A.,   Seibert S.,  2015, in Proceedings of the Second
  Workshop on the LLVM Compiler Infrastructure in HPC. LLVM '15.
ACM, New York, NY, USA, pp 7:1--7:6, \mn@doi{10.1145/2833157.2833162}, \url
  {http://doi.acm.org/10.1145/2833157.2833162}

\bibitem[\protect\citeauthoryear{{Laplace}}{{Laplace}}{1774}]{laplace}
{Laplace} P.,  1774, l'Academie Royale des Sciences, 6, 621

\bibitem[\protect\citeauthoryear{Lindsay \& Liu}{Lindsay \&
  Liu}{2009}]{lindsay2009}
Lindsay B.,  Liu J.,  2009, \mn@doi [Statist. Sci.] {10.1214/09-STS302}, 24,
  303

\bibitem[\protect\citeauthoryear{{Lisakov}, {Kovalev}, {Savolainen}, {Hovatta}
  \& {Kutkin}}{{Lisakov} et~al.}{2017}]{2017MNRAS.468.4478L}
{Lisakov} M.~M.,  {Kovalev} Y.~Y.,  {Savolainen} T.,  {Hovatta} T.,   {Kutkin}
  A.~M.,  2017, \mn@doi [\mnras] {10.1093/mnras/stx710}, \href
  {http://adsabs.harvard.edu/abs/2017MNRAS.468.4478L} {468, 4478}

\bibitem[\protect\citeauthoryear{{Lister} et~al.,}{{Lister}
  et~al.}{2009a}]{2009AJ....137.3718L}
{Lister} M.~L.,  et~al., 2009a, \mn@doi [\aj] {10.1088/0004-6256/137/3/3718},
  \href {http://adsabs.harvard.edu/abs/2009AJ....137.3718L} {137, 3718}

\bibitem[\protect\citeauthoryear{{Lister} et~al.,}{{Lister}
  et~al.}{2009b}]{mojave6kinematicsanalysis}
{Lister} M.~L.,  et~al., 2009b, \mn@doi [\aj] {10.1088/0004-6256/138/6/1874},
  \href {http://adsabs.harvard.edu/abs/2009AJ....138.1874L} {138, 1874}

\bibitem[\protect\citeauthoryear{{Lobanov}}{{Lobanov}}{1998}]{1998A&A...330...79L}
{Lobanov} A.~P.,  1998, \aap, \href
  {http://adsabs.harvard.edu/abs/1998A%26A...330...79L} {330, 79}

\bibitem[\protect\citeauthoryear{{Lobanov} \& {Zensus}}{{Lobanov} \&
  {Zensus}}{1999}]{1999ApJ...521..509L}
{Lobanov} A.~P.,  {Zensus} J.~A.,  1999, \mn@doi [\apj] {10.1086/307555}, \href
  {http://adsabs.harvard.edu/abs/1999ApJ...521..509L} {521, 509}

\bibitem[\protect\citeauthoryear{{Lyutikov}, {Pariev}  \& {Gabuzda}}{{Lyutikov}
  et~al.}{2005}]{2005MNRAS.360..869L}
{Lyutikov} M.,  {Pariev} V.~I.,   {Gabuzda} D.~C.,  2005, \mn@doi [\mnras]
  {10.1111/j.1365-2966.2005.08954.x}, \href
  {http://adsabs.harvard.edu/abs/2005MNRAS.360..869L} {360, 869}

\bibitem[\protect\citeauthoryear{{Malyshev}, {Zdziarski}  \&
  {Chernyakova}}{{Malyshev} et~al.}{2013}]{2013MNRAS.434.2380M}
{Malyshev} D.,  {Zdziarski} A.~A.,   {Chernyakova} M.,  2013, \mn@doi [\mnras]
  {10.1093/mnras/stt1184}, \href
  {http://adsabs.harvard.edu/abs/2013MNRAS.434.2380M} {434, 2380}

\bibitem[\protect\citeauthoryear{{Marcaide} \& {Shapiro}}{{Marcaide} \&
  {Shapiro}}{1984}]{1984ApJ...276...56M}
{Marcaide} J.~M.,  {Shapiro} I.~I.,  1984, \mn@doi [\apj] {10.1086/161592},
  \href {http://adsabs.harvard.edu/abs/1984ApJ...276...56M} {276, 56}

\bibitem[\protect\citeauthoryear{{Marscher}}{{Marscher}}{2009}]{2009ASPC..402..194M}
{Marscher} A.~P.,  2009, in {Hagiwara} Y.,  {Fomalont} E.,  {Tsuboi} M.,
  {Yasuhiro} M.,  eds,  Astronomical Society of the Pacific Conference Series
  Vol. 402, Approaching Micro-Arcsecond Resolution with VSOP-2: Astrophysics
  and Technologies. p.~194

\bibitem[\protect\citeauthoryear{{Morganti}, {Peck}, {Oosterloo}, {van
  Moorsel}, {Capetti}, {Fanti}, {Parma}  \& {de Ruiter}}{{Morganti}
  et~al.}{2009}]{2009A&A...505..559M}
{Morganti} R.,  {Peck} A.~B.,  {Oosterloo} T.~A.,  {van Moorsel} G.,  {Capetti}
  A.,  {Fanti} R.,  {Parma} P.,   {de Ruiter} H.~R.,  2009, \mn@doi [\aap]
  {10.1051/0004-6361/200912605}, \href
  {https://ui.adsabs.harvard.edu/\#abs/2009A&A...505..559M} {505, 559}

\bibitem[\protect\citeauthoryear{{Natarajan}, {Paragi}, {Zwart}, {Perkins},
  {Smirnov}  \& {van der Heyden}}{{Natarajan}
  et~al.}{2017}]{2017MNRAS.464.4306N}
{Natarajan} I.,  {Paragi} Z.,  {Zwart} J.,  {Perkins} S.,  {Smirnov} O.,   {van
  der Heyden} K.,  2017, \mn@doi [\mnras] {10.1093/mnras/stw2653}, \href
  {http://adsabs.harvard.edu/abs/2017MNRAS.464.4306N} {464, 4306}

\bibitem[\protect\citeauthoryear{{Nokhrina}}{{Nokhrina}}{2017}]{2017MNRAS.468.2372N}
{Nokhrina} E.~E.,  2017, \mn@doi [\mnras] {10.1093/mnras/stx521}, \href
  {http://adsabs.harvard.edu/abs/2017MNRAS.468.2372N} {468, 2372}

\bibitem[\protect\citeauthoryear{{O'Sullivan} \& {Gabuzda}}{{O'Sullivan} \&
  {Gabuzda}}{2009}]{2009MNRAS.400...26O}
{O'Sullivan} S.~P.,  {Gabuzda} D.~C.,  2009, \mn@doi [\mnras]
  {10.1111/j.1365-2966.2009.15428.x}, \href
  {http://adsabs.harvard.edu/abs/2009MNRAS.400...26O} {400, 26}

\bibitem[\protect\citeauthoryear{Pjanka, Zdziarski  \& Sikora}{Pjanka
  et~al.}{2016}]{10.1093/mnras/stw2960}
Pjanka P.,  Zdziarski A.~A.,   Sikora M.,  2016, \mn@doi [Monthly Notices of
  the Royal Astronomical Society] {10.1093/mnras/stw2960}, 465, 3506

\bibitem[\protect\citeauthoryear{{Plavin}, {Kovalev}, {Pushkarev}  \&
  {Lobanov}}{{Plavin} et~al.}{2019}]{2019MNRAS.485.1822P}
{Plavin} A.~V.,  {Kovalev} Y.~Y.,  {Pushkarev} A.~B.,   {Lobanov} A.~P.,  2019,
  \mn@doi [\mnras] {10.1093/mnras/stz504}, \href
  {https://ui.adsabs.harvard.edu/abs/2019MNRAS.485.1822P} {485, 1822}

\bibitem[\protect\citeauthoryear{{Potter} \& {Cotter}}{{Potter} \&
  {Cotter}}{2013}]{2013MNRAS.429.1189P}
{Potter} W.~J.,  {Cotter} G.,  2013, \mn@doi [\mnras] {10.1093/mnras/sts407},
  \href {http://adsabs.harvard.edu/abs/2013MNRAS.429.1189P} {429, 1189}

\bibitem[\protect\citeauthoryear{{Pushkarev}, {Hovatta}, {Kovalev}, {Lister},
  {Lobanov}, {Savolainen}  \& {Zensus}}{{Pushkarev}
  et~al.}{2012}]{2012A&A...545A.113P}
{Pushkarev} A.~B.,  {Hovatta} T.,  {Kovalev} Y.~Y.,  {Lister} M.~L.,  {Lobanov}
  A.~P.,  {Savolainen} T.,   {Zensus} J.~A.,  2012, \mn@doi [\aap]
  {10.1051/0004-6361/201219173}, \href
  {http://adsabs.harvard.edu/abs/2012A%26A...545A.113P} {545, A113}

\bibitem[\protect\citeauthoryear{{Pushkarev}, {Kovalev}, {Lister}  \&
  {Savolainen}}{{Pushkarev} et~al.}{2017}]{2017MNRAS.468.4992P}
{Pushkarev} A.~B.,  {Kovalev} Y.~Y.,  {Lister} M.~L.,   {Savolainen} T.,  2017,
  \mn@doi [\mnras] {10.1093/mnras/stx854}, \href
  {https://ui.adsabs.harvard.edu/#abs/2017MNRAS.468.4992P} {468, 4992}

\bibitem[\protect\citeauthoryear{{Reynolds}, {Fabian}, {Celotti}  \&
  {Rees}}{{Reynolds} et~al.}{1996}]{1996MNRAS.283..873R}
{Reynolds} C.~S.,  {Fabian} A.~C.,  {Celotti} A.,   {Rees} M.~J.,  1996,
  \mn@doi [\mnras] {10.1093/mnras/283.3.873}, \href
  {http://adsabs.harvard.edu/abs/1996MNRAS.283..873R} {283, 873}

\bibitem[\protect\citeauthoryear{{Richards} et~al.,}{{Richards}
  et~al.}{2011}]{2011ApJS..194...29R}
{Richards} J.~L.,  et~al., 2011, \mn@doi [The Astrophysical Journal Supplement
  Series] {10.1088/0067-0049/194/2/29}, \href
  {https://ui.adsabs.harvard.edu/\#abs/2011ApJS..194...29R} {194, 29}

\bibitem[\protect\citeauthoryear{{Rybicki} \& {Lightman}}{{Rybicki} \&
  {Lightman}}{1979}]{1979rpa..book.....R}
{Rybicki} G.~B.,  {Lightman} A.~P.,  1979, {Radiative processes in
  astrophysics}

\bibitem[\protect\citeauthoryear{{Shepherd}}{{Shepherd}}{1997}]{1997ASPC..125...77S}
{Shepherd} M.~C.,  1997, in {G.~Hunt \& H.~Payne} ed.,  Astronomical Society of
  the Pacific Conference Series Vol. 125, Astronomical Data Analysis Software
  and Systems VI. p.~77

\bibitem[\protect\citeauthoryear{{Sironi} \& {Spitkovsky}}{{Sironi} \&
  {Spitkovsky}}{2014}]{2014ApJ...783L..21S}
{Sironi} L.,  {Spitkovsky} A.,  2014, \mn@doi [\apjl]
  {10.1088/2041-8205/783/1/L21}, \href
  {http://adsabs.harvard.edu/abs/2014ApJ...783L..21S} {783, L21}

\bibitem[\protect\citeauthoryear{{Sironi}, {Spitkovsky}  \& {Arons}}{{Sironi}
  et~al.}{2013}]{2013ApJ...771...54S}
{Sironi} L.,  {Spitkovsky} A.,   {Arons} J.,  2013, \mn@doi [\apj]
  {10.1088/0004-637X/771/1/54}, \href
  {https://ui.adsabs.harvard.edu/#abs/2013ApJ...771...54S} {771, 54}

\bibitem[\protect\citeauthoryear{{Sironi}, {Keshet}  \& {Lemoine}}{{Sironi}
  et~al.}{2015}]{2015SSRv..191..519S}
{Sironi} L.,  {Keshet} U.,   {Lemoine} M.,  2015, \mn@doi [\ssr]
  {10.1007/s11214-015-0181-8}, \href
  {http://adsabs.harvard.edu/abs/2015SSRv..191..519S} {191, 519}

\bibitem[\protect\citeauthoryear{Skilling}{Skilling}{2004}]{doi:10.1063/1.1835238}
Skilling J.,  2004, \mn@doi [AIP Conference Proceedings] {10.1063/1.1835238},
  735, 395

\bibitem[\protect\citeauthoryear{{Sokolovsky}, {Kovalev}, {Pushkarev}  \&
  {Lobanov}}{{Sokolovsky} et~al.}{2011}]{sokolovsky}
{Sokolovsky} K.~V.,  {Kovalev} Y.~Y.,  {Pushkarev} A.~B.,   {Lobanov} A.~P.,
  2011, \mn@doi [\aap] {10.1051/0004-6361/201016072}, \href
  {http://adsabs.harvard.edu/abs/2011A%26A...532A..38S} {532, A38}

\bibitem[\protect\citeauthoryear{{Thompson}, {Moran}  \& {Swenson}}{{Thompson}
  et~al.}{2017}]{2017isra.book.....T}
{Thompson} A.~R.,  {Moran} J.~M.,   {Swenson} Jr. G.~W.,  2017, {Interferometry
  and Synthesis in Radio Astronomy, 3rd Edition},
  \mn@doi{10.1007/978-3-319-44431-4.
}

\bibitem[\protect\citeauthoryear{{Urry} \& {Padovani}}{{Urry} \&
  {Padovani}}{1995}]{1995PASP..107..803U}
{Urry} C.~M.,  {Padovani} P.,  1995, \mn@doi [\pasp] {10.1086/133630}, \href
  {http://adsabs.harvard.edu/abs/1995PASP..107..803U} {107, 803}

\bibitem[\protect\citeauthoryear{Van Der~Walt, Colbert  \& Varoquaux}{Van
  Der~Walt et~al.}{2011}]{numpy}
Van Der~Walt S.,  Colbert S.~C.,   Varoquaux G.,  2011, \mn@doi [Computing in
  Science \\& Engineering] {10.1109/MCSE.2011.37}, 13, 22

\bibitem[\protect\citeauthoryear{{Venturi}, {Giovannini}, {Feretti},
  {Comoretto}  \& {Wehrle}}{{Venturi} et~al.}{1993}]{1993ApJ...408...81V}
{Venturi} T.,  {Giovannini} G.,  {Feretti} L.,  {Comoretto} G.,   {Wehrle}
  A.~E.,  1993, \mn@doi [\apj] {10.1086/172571}, \href
  {http://adsabs.harvard.edu/abs/1993ApJ...408...81V} {408, 81}

\bibitem[\protect\citeauthoryear{{Werner} \& {Uzdensky}}{{Werner} \&
  {Uzdensky}}{2017}]{2017ApJ...843L..27W}
{Werner} G.~R.,  {Uzdensky} D.~A.,  2017, \mn@doi [\apjl]
  {10.3847/2041-8213/aa7892}, \href
  {http://adsabs.harvard.edu/abs/2017ApJ...843L..27W} {843, L27}

\bibitem[\protect\citeauthoryear{{Werner}, {Uzdensky}, {Begelman}, {Cerutti}
  \& {Nalewajko}}{{Werner} et~al.}{2018}]{2018MNRAS.473.4840W}
{Werner} G.~R.,  {Uzdensky} D.~A.,  {Begelman} M.~C.,  {Cerutti} B.,
  {Nalewajko} K.,  2018, \mn@doi [\mnras] {10.1093/mnras/stx2530}, \href
  {http://adsabs.harvard.edu/abs/2018MNRAS.473.4840W} {473, 4840}

\bibitem[\protect\citeauthoryear{{Worrall}, {Birkinshaw}, {Laing}, {Cotton}  \&
  {Bridle}}{{Worrall} et~al.}{2007}]{2007MNRAS.380....2W}
{Worrall} D.~M.,  {Birkinshaw} M.,  {Laing} R.~A.,  {Cotton} W.~D.,   {Bridle}
  A.~H.,  2007, \mn@doi [\mnras] {10.1111/j.1365-2966.2007.11998.x}, \href
  {http://adsabs.harvard.edu/abs/2007MNRAS.380....2W} {380, 2}

\bibitem[\protect\citeauthoryear{Zdziarski}{Zdziarski}{2014}]{10.1093/mnras/stu1835}
Zdziarski A.~A.,  2014, \mn@doi [Monthly Notices of the Royal Astronomical
  Society] {10.1093/mnras/stu1835}, 445, 1321

\bibitem[\protect\citeauthoryear{{Zdziarski}, {Sikora}, {Pjanka}  \&
  {Tchekhovskoy}}{{Zdziarski} et~al.}{2015}]{2015MNRAS.451..927Z}
{Zdziarski} A.~A.,  {Sikora} M.,  {Pjanka} P.,   {Tchekhovskoy} A.,  2015,
  \mn@doi [\mnras] {10.1093/mnras/stv986}, \href
  {http://adsabs.harvard.edu/abs/2015MNRAS.451..927Z} {451, 927}

\makeatother
\end{thebibliography}




\appendix

\clearpage

\section{Deriving original formulas}
\label{sec:deriveBK}

Here we derive the essential formula from the original paper \citep{1979ApJ...232...34B} for completeness.
Coefficients of emission and absorption for for synchrotron radiation in the observer frame expressed in terms of the plasma rest frame values of $B$ and $K$ in case of tangled magnetic field, power law energy distribution of the emitting particles and isotropic pitch-angles distribution \citep{1979rpa..book.....R}: 
\begin{equation}
\label{eq:eta}
\eta(\nu_{\mathrm{obs}}) = (4\pi)^{-1}\langle C_{\rm 1}(\alpha)\rangle(1+z)^{-\alpha}D^{2+\alpha}KB^{1+\alpha}\nu_{\mathrm{obs}}^{-\alpha}
\end{equation}  
%
\begin{equation}
\label{eq:kappa}
\kappa(\nu_{\mathrm{obs}}) = \langle C_{\rm 2}(\alpha)\rangle(1+z)^{-(\alpha+2.5)}D^{1.5+\alpha}KB^{1.5+\alpha}\nu_{\mathrm{obs}}^{-(2.5+\alpha)}
\end{equation}
where $z$ is the redshift, $\alpha$ is optically thin spectral index related to the emitting particles powe-law exponent $\alpha = (s-1)/2$, $D$ -- Doppler factor, $\langle C_1(\alpha)\rangle $ and $\langle C_2(\alpha)\rangle$ are averaged over isotropic pitch-angle distributions\footnote{For tangled magnetic field one has to integrate $C_1(\alpha)(\sin{\theta})^{\alpha+1}$ and $C_2(\alpha)(\sin{\theta})^{\alpha+1.5}$ over all directions} values of the coefficients, expressed through \texttt{Gamma} functions as follows:
\begin{equation}
\label{eq:c1averaged}
    \langle C_1(\alpha) \rangle = \sqrt{\frac{\pi}{4}} C_1(\alpha)\frac{\Gamma\left(\frac{3+\alpha}{2}\right)}{\Gamma\left(\frac{4+\alpha}{2}\right)}
\end{equation}
\begin{equation}
\label{eq:c2averaged}
    \langle C_2(\alpha) \rangle = \sqrt{\frac{\pi}{4}} C_2(\alpha)\frac{\Gamma\left(\frac{7+2\alpha}{4}\right)}{\Gamma\left(\frac{9+2\alpha}{4}\right)}
\end{equation}
where
\begin{equation}
\label{eq:c1}
    C_1(\alpha) = \frac{\sqrt{3}e^3}{m c^2(2\alpha+2)}\left(\frac{3e}{2\pi m c}\right)^{\alpha}\Gamma\left(\frac{\alpha}{2}+\frac{11}{6}\right)\Gamma\left(\frac{\alpha}{2}+\frac{1}{6}\right)
\end{equation}

\begin{equation}
\label{eq:c2}
    C_2(\alpha) = \frac{\sqrt{3}e^3}{8\pi m^2 c^2}\left(\frac{3e}{2\pi m c}\right)^{\alpha+0.5}\Gamma\left(\frac{6\alpha+5}{12}\right)\Gamma\left(\frac{6\alpha+25}{12}\right)
\end{equation}

Using (\ref{eq:kappa}) and expressing geometrical length $L$ along line of sight at a map point parametrized by the observed distances along the jet $r_{\rm obs} = r \sin{\theta}$ and in perpendicular direction $d$ as:

\begin{equation}
\label{eq:geompath}
    L = 2 r \sqrt{\phi^2-\left(\frac{d}{r}\right)^2}/ \sin{\theta}  
\end{equation}
where $\phi$ is half-opening angle of the cone and $\theta$ -- jet viewing angle, one can derive the expression for optical depth $\tau= \int \kappa dl$. The last can be approximated as $\tau = \kappa L$, where $\kappa$ is the value of the absorption coefficient at the apex of the jet for a given line of sight:
\begin{multline}
\label{eq:tauphiapp}
\tau(r_{\rm obs}, d) =  C_{\rm 2}(\alpha)(1+z)^{-(\alpha+2.5)}D^{1.5+\alpha}KB^{1.5+\alpha}\nu_{\mathrm{obs}}^{-(2.5+\alpha)} \\
\times 2 r \sqrt{\phi^2-\left(\frac{d}{r}\right)^2}/ \sin{\theta}  
\end{multline}
Here $B$, $K$ and $D$ are values at point $r = r_{\rm obs}/\sin{\theta}$, $C_1(\alpha)$ and $C_2(\alpha)$ are already averaged over isotropic direction of the tangled magnetic field coefficients, i.e. it is (\ref{eq:c1averaged}) and (\ref{eq:c2averaged}).
Substituting values of fields $B$ and $K$ at point $r$ (i.e. $B = B_1 (r/r_1)^{-1}$, where $r_1$ is 1 pc, and $K = K_1 (r/r_1)^{-2}$) and assuming $\alpha=0.5$, $m=1$, $n=2$:
\begin{multline}
\label{eq:tauphiapp1}
\tau(r_{\rm obs}, d) =  C_{\rm 2}(\alpha)(1+z)^{-3}D^{2}K_1 B_1^{2}\nu_{\mathrm{obs}}^{-3} \\
\times \left(\frac{r}{r_1}\right)^{-4}  2 r \sqrt{\phi^2-\left(\frac{d}{r}\right)^2}/ \sin{\theta}  
\end{multline}
where $r_1$ is 1 pc.
Substituting $r_{\rm ob}$ instead of $r$:
\begin{multline}
\label{eq:tauphiapp2}
\tau(r_{\rm obs}, d) =  C_{\rm 2}(\alpha)(1+z)^{-3}D^{2}K_1 B_1^{2}\nu_{\mathrm{obs}}^{-3} \\
\times 2r^4_{1}(\sin{\theta})^2 \sqrt{\phi^2-\left(\frac{d}{r}\right)^2}r_{\rm ob}^{-3}
\end{multline}
This coincides with (26) from \cite{1979ApJ...232...34B}. Using \textit{cgs} values of the constants\footnote{These are already averaged values, i.e. it is (\ref{eq:c1averaged}) and (\ref{eq:c2averaged})} $C_1(0.5)=3.6\cdot 10^{-19}$ and $C_2(0.5)=9.24\cdot 10^{10}$ we obtain coefficient in (\ref{eq:tauphiapp2}) $\approx570.4$. That is close to the original 500 in equation (26) in \citep{1979ApJ...232...34B}. To completely move from true half-opening angle and distance from cone apex ($\phi$, $r$) to the observed values ($\phi_{\rm app}$, $r_{\rm obs}$) getting $\sin{\theta}$ out of the brackets gives:
\begin{multline}
\label{eq:tauphiapp3}
\tau(r_{\rm obs}, d) =  C_{\rm 2}(\alpha)(1+z)^{-3}D^{2}K_1 B_1^{2}\nu_{\mathrm{obs}}^{-3} \\
\times 2r^4_{1}(\sin{\theta})^3 \sqrt{\phi_{\rm app}^2-\left(\frac{d}{r_{\rm obs}}\right)^2}r_{\rm obs}^{-3}
\end{multline}

For brightness distribution using the solution of the radiative transfer equation:
\begin{equation}
\label{eq:transfersolution}
    I_{\nu}(\nu_{\mathrm{obs}}) = \frac{\eta(\nu_{\mathrm{obs}})}{\kappa(\nu_{\mathrm{obs}})}(1 - e^{-\tau})
\end{equation}
expressions for coefficients (\ref{eq:eta})--(\ref{eq:kappa}) and Lorentz invariant $\frac{I_{\nu}}{\nu^3}$ we obtain:
\begin{multline}
    I_{\nu}(r_{\rm obs}, d) = \frac{C_{1}(\alpha)D^{0.5}\nu^{2.5}_{\rm obs}}{4\pi C_2(\alpha) (1+z)^{0.5} B^{0.5}}(1 - e^{-\tau(r_{\rm obs}, d)})
\end{multline}
To compare with \cite{1979ApJ...232...34B} results we express the brightness temperature:
\begin{equation}
\label{eq:Tbours1}
T(r_{\rm obs}, d) = \frac{c^2 C_{1}(\alpha)D^{0.5}\nu^{0.5}_{\rm obs}}{8\pi k_B C_2(\alpha) (1+z)^{0.5} B^{0.5}}(1 - e^{-\tau(r_{\rm obs}, d)})
\end{equation}
Substituting the magnetic field coordinate dependence and using $r_{\rm obs}$ instead of $r$ we obtain:
\begin{equation}
\label{eq:Tbours2}
T(r_{\rm obs}, d) = \frac{c^2 C_{1}(\alpha)D^{0.5}\nu^{0.5}_{\rm obs}r_{\rm obs}^{0.5}}{8\pi k_B C_2(\alpha) (1+z)^{0.5} B_1^{0.5}r_1^{0.5}(\sin{\theta})^{0.5}}(1 - e^{-\tau(r_{\rm obs}, d)})
\end{equation}
where $r_1$ is 1 pc.

Using \textit{cgs} values of the constants $C_1(0.5)=3.6\cdot 10^{-19}$ and $C_2(0.5)=9.24\cdot 10^{10}$ we obtain coefficient $\approx3.186\cdot10^{10}$K that coincides with that from equation (25) in \cite{1979ApJ...232...34B}.

\section{Extending for arbitrary $\alpha$, \texorpdfstring{\MakeLowercase{$m$}}{$m$}, \texorpdfstring{\MakeLowercase{$n$}}{$n$}}
\label{sec:extendingBK}

Generalization to arbitrary $m$, $n$, $\alpha$ is obvious. One should use $(r/r_1)^{-m}$ instead of $(r/r_1)^{-1}$ (similar for $K$) when going from (\ref{eq:tauphiapp}) to (\ref{eq:tauphiapp1}) and from (\ref{eq:Tbours1}) to (\ref{eq:Tbours2}) and keep $\alpha$ after (\ref{eq:tauphiapp}). For the optical depth we obtain:
\begin{multline}
\label{eq:tauphiapp4}
\tau(r_{\rm obs}, d) =  C_{\rm 2}(\alpha)(1+z)^{-(\alpha+2.5)}D^{1.5+\alpha}K_1 B_1^{1.5+\alpha}\nu_{\mathrm{obs}}^{-(2.5+\alpha)} \\
\times 2r^{n+m(1.5+\alpha)}_{1}\left(\sin{\theta}\right)^{n+m(1.5+\alpha)-1} \sqrt{\phi_{\rm app}^2-\left(\frac{d}{r_{\rm obs}}\right)^2}r_{\rm ob}^{-(n+m(1.5+\alpha)-1)}
\end{multline}
For canonical $n=2$, $m=1$, $\alpha=0.5$ we obtain $\tau(r_{\rm obs}, d) \propto r^{-3}_{\rm obs}$ in accordance with (\ref{eq:tauphiapp3}). The numerical coefficient strongly depends on $\alpha$ through $C_2(\alpha)$. E.g. for $\alpha=0.55$ it is $\approx1237$ and for $\alpha=0.75$ it is $\approx27980$.

For brightness temperature:
\begin{equation}
\label{eq:Tbours3}
T(r_{\rm obs}, d) = \frac{c^2 C_{1}(\alpha)D^{0.5}\nu^{0.5}_{\rm obs}r_{\rm obs}^{0.5m}}{8\pi k_B C_2(\alpha) (1+z)^{0.5} B_1^{0.5}r_1^{0.5m}(\sin{\theta})^{0.5m}}(1 - e^{-\tau(r_{\rm obs}, d)})
\end{equation}
where $\tau(r_{\rm obs}, d)$ is from (\ref{eq:tauphiapp4}) and numerical coefficient depends only on $\alpha$. E.g. for $\alpha=0.75$ it is $\approx2.37\cdot10^{10}$.

To find position of the core (i.e. maximum intensity) at some frequency one has to differentiate Eq.~\ref{eq:Tbours3} with respect to $r_{\rm obs}$. This way the following relation for the optical depth at the maximum intensity is obtained:
$$\tau = \log(1+\frac{n+m(1.5+\alpha)-1}{0.5m}\tau)$$
For $n=2$, $m=1$ we obtain $\tau \approx 2.92$. Thus to obtain position of maximum $r_{\rm max}$ for some frequency one has to solve equation Eq.~\ref{eq:tauphiapp4} for $r_{\rm obs}$ given $\tau = 2.92$,

\section{``Natural'' parametrization}
\label{sec:naturalparametrization}

With (\ref{eq:tauphiapp4}) and (\ref{eq:Tbours3}) the essential parametrization:
\begin{equation}
\label{eq:a1natural}
    A_1 = \left(\frac{D}{B_1(\sin{\theta})^m}\right)^{0.5}
\end{equation}
\begin{equation}
\label{eq:a2natural}
    A_2 = K_1 (D B_1)^{1.5+\alpha}(1+z)^{-(\alpha+2.5)}(\sin{\theta})^{n+m(1.5+\alpha)-1}
\end{equation}

With this parametrization:
\begin{equation}
\label{eq:TparametrizationNatural}
T = C_T(\alpha) A_1 \left(\frac{\nu_{\rm obs} r_\text{obs}^{m}}{(1 + z)}\right)^{1/2}  (1 - e^{-\tau})
\end{equation}
\begin{equation}
\label{eq:tauparametrizationNatural}
\tau =C_{\tau}(\alpha)  A_2 \nu_{\rm obs}^{-(2.5+\alpha)} \sqrt{\phi_\text{app}^2 - \left(\frac{d}{r_{\rm obs}}\right)^2} r_{\rm ob}^{-(n+m(1.5+\alpha)-1)}
\end{equation}
where the observed frequency $\nu_{\rm obs}$ is in GHz and projected distance from cone apex $r_{\rm obs}$ is in pc and $C_T(\alpha$) and $C_{\tau}(\alpha)$ are:
\begin{equation}
\label{eq:Ctau}
    C_{\tau}(\alpha) = 2\langle C_2(\alpha) \rangle
    \left(\frac{\rm GHz}{\rm Hz}\right)^{-2.5-\alpha}
    (1 \rm pc)^{n+m(\alpha+1.5)}
    \left(\frac{\rm pc}{\rm cm}\right)^{-n-m(\alpha+1.5)+1}
\end{equation}
\begin{equation}
\label{eq:CT}
    C_{T}(\alpha) = \frac{c^2 \langle C_1(\alpha) \rangle
    \left(\frac{\rm GHz}{\rm Hz}\right)^{0.5}
    \left(\frac{\rm pc}{\rm cm}\right)^{0.5m}}{8\pi k_{\rm B} \langle C_2(\alpha) \rangle (1 \rm pc)^{0.5m}}
\end{equation}
expressed in \textit{cgs} units with values that depend on $\alpha$ only. For example, $C_{\tau}(0.5) \approx 570.4$ in accordance with that from (\ref{eq:tauphiapp2}) and $C_{T}(0.5) \approx 3.186 \cdot 10^{10}$ in accordance with formula (26) and (25) from \citep{1979ApJ...232...34B}.

Recall that for intrinsically symmetrical counter jet the parameters of are related to that of the approaching jet:
$$A_1^{\rm cj} = \frac{A_1}{\sqrt{R}}$$
$$A_2^{\rm cj} = \frac{A_2}{R^{1.5+\alpha}}$$
where $R = (1+\beta \cos{\theta})/(1-\beta \cos{\theta})$ is the ratio of the Doppler factors of the approaching and receding jets.

Physical parameters can be obtained in ``natural'' parametrization as:
\begin{equation}
    B_1 = \frac{D}{A_1^2 (\sin{\theta})^m} 
\end{equation}
\begin{equation}
    K_1 = \frac{A_2 A_1^{3+2\alpha}}{D^{3+2\alpha} (\sin{\theta})^{n-1} (1+z)^{-(\alpha+2.5)}} 
\end{equation}
\begin{equation}
    \frac{B_1^2}{K_1} = \frac{D^{5+2\alpha}(1+z)^{-(\alpha+2.5)}(\sin{\theta})^{n-2m-1}}{A_1^{7+2\alpha}A_2} 
\end{equation}

\section{Accounting for the uncertainty in the amplitude scale factors}
\label{sec:accountmeangain}

After sampling of the posterior distribution with the constrained ($\prod_b g_{b} = 1$) amplitude scale factors $g_b$ we can infer the uncertainty resulting from their product not being exactly 1. As we noted in Section~\ref{sec:reparametrization}, multiplication of all scale factors by the same number is equivalent to multipliction of $A_1$ on this number. Thus, corresponding uncertainty of the $A_1$ is the uncertainty of the mean of the scale factors. We used priors $N(1, \sigma_b)$ for $g_b$ at each frequency band, where $\sigma_b$ -- absolute calibration uncertainty estimated by \cite{2014AJ....147..143H} and samples from the obtained posteriors of $A_1$, $A_2$:
\begin{equation}
    A_1^* = A_1 N(1, \sigma_{\rm gains}/n_{b})\\
    A_2^* = A_2
\label{eq:accountmeangainnatural}
\end{equation}
where $\sigma_{\rm gains}/n_{b}$ -- uncertainty of the mean of scale factors deduced from their priors and $n_b$ -- number of frequency bands.



\bsp	
\label{lastpage}
\end{document}